\documentclass[12pt,preprint]{aastex}
\usepackage{natbib}
\usepackage{lscape}
\usepackage{graphicx}
\usepackage{amssymb}
\usepackage{epsfig}
\usepackage{amsmath}
\usepackage{graphicx}

\newcommand       \msun        	{$M_{\odot}$}
\newcommand       \lsun      	{$L_{\odot}$} 

\newcommand	     \cc             {cm$^{-3}$}

\newcommand       \mic        	 {$\mu$m}

\newcommand	     \cosae              {\cos(\alpha)}
\newcommand 		\spitz		{{\it Spitzer}}
\newcommand		\iras			{{\it IRAS}}
\newcommand		\lburst			{$L_b$}

\newcommand		\taub			{$\Delta t_b$}

\newcommand		\euva			{EUV18}
\newcommand		\euvb			{EUV19}
\newcommand		\euvc			{EUV20}
\newcommand		\uva		{UV18}
\newcommand		\uvb		{UV19}
\newcommand		\uvc			{UV20}
\newcommand		\olc		{OPT}
\newcommand		\xsi			{$\xi_F$}
\newcommand		\chisq      {$\chi^2$}

\begin{document}

\title{INFRARED ECHOES REVEAL THE SHOCK BREAKOUT \\OF THE CAS A SUPERNOVA}

\author{Eli Dwek\altaffilmark{1} and Richard G. Arendt \altaffilmark{2}}
\altaffiltext{1}{Observational Cosmology Lab., Code 665; NASA Goddard
Space Flight Center, Greenbelt, MD 20771, U.S.A., e-mail: eli.dwek@nasa.gov}
\altaffiltext{2}{University of Maryland Baltimore County (UMBC), Code 665, NASA Goddard Space Flight Center, Greenbelt MD, 20771, U.S.A., e-mail: richard.g.arendt@nasa.gov}  

\begin{abstract}
The serendipitous discovery of infrared echoes around the Cas~A supernova remnant the \spitz\ satellite has provided astronomers with a unique opportunity to study the properties of the echoing material and the history and nature of the outburst that generated these echoes. In retrospect, we find that the echoes are also clearly visible as infrared ``hot spots" in \iras\ images of the region.  
The spectra of the echoes are distinct from that of the dust in the general diffuse interstellar medium (ISM) revealing hot silicate grains that are either stochastically heated to temperatures in excess of $\sim 150$~K, or radiating at an equilibrium temperature of this value. We show that the maximum  luminosity that can be generated by the radioactive decay of $^{56}$Ni is not capable of producing such spectra, and could therefore not have given rise to the echoes.  
Instead, we find that the echoes must have been generated by an intense and short burst of EUV-UV radiation  associated with the breakout of the shock through the surface of the exploding star. The inferred luminosity of the burst depends on the amount of attenuation in the intervening medium to the clouds, and we derive a burst luminosity of $\sim 1.5\times10^{11}$~\lsun\ for an assumed H-column density of $1.5\times10^{19}$~cm$^{-2}$. The average H-column density of the IR emitting region in the echoing clouds is about $5\times10^{17}$~cm$^{-2}$. Derivation of their density requires knowledge of the width of the echo that is sweeping though the ISM, which in turn is determined by the duration of the burst. A burst time of $\sim 1$~d gives a cloud density of $\sim 200$~\cc, typical of dense IR cirrus. 
\end{abstract}

\section{INTRODUCTION}
Short-lived luminous sources can produce echoes of their outburst in the ISM. These echoes can be manifested  as line emission from the gas, or reflected and thermally-reradiated light from the dust in the ISM. They can be used to probe the morphology of the interstellar medium (ISM) through which they are expanding and to reconstruct the historical record of the temporal behavior of the light source.  

Visible light echoes have been 
detected around Nova Persei 1901 \citep{kapteyn02,bode04}, V838 Mon \citep{bond03}, and SN 1987A \citep{crotts88}. Fluorescent echoes arising from the equatorial ring around SN1987A have been used to probe the early hours of the burst from the SN, the radius of the ring, and the distance to the LMC \citep{fransson89b, panagia91, dwek92d}.
The SuperMACHO project has discovered three additional sets
of light echoes in the LMC which can be traced back to known supernova remnants (SNRs), and were used to derive ages of 610 and 410 yr for two of the SNRs \citep{rest05}. 

\begin{figure}[htbp]
  \centering
  \includegraphics[width=5.0in]{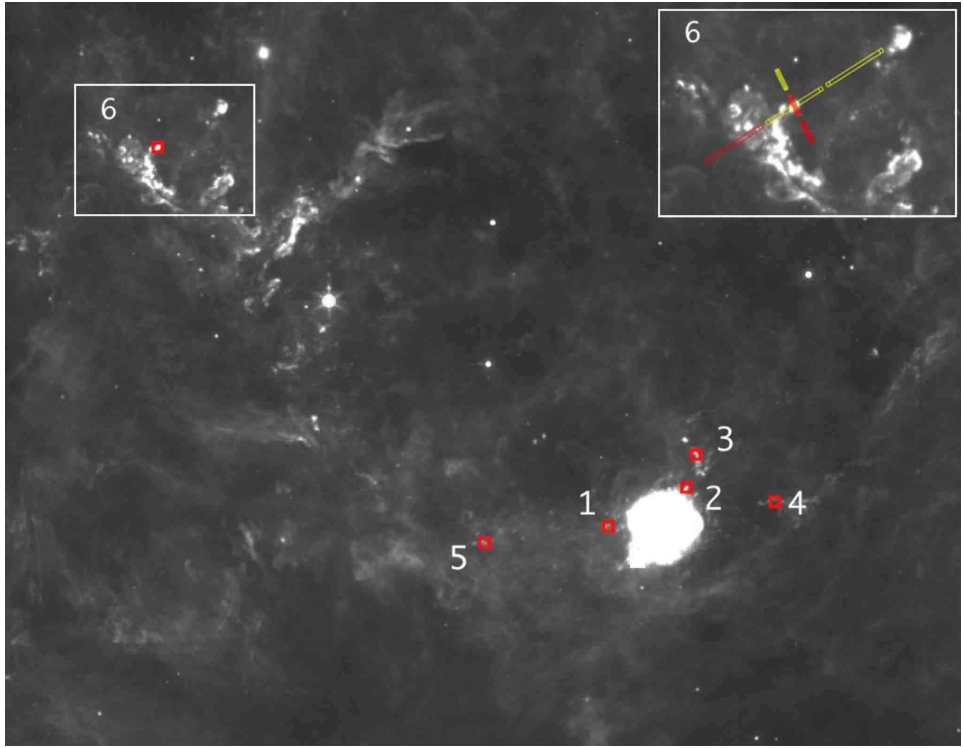}
  \caption{{\footnotesize
A map of the region around Cas A showing  the location of all transient light signals identified as echoes (red squares). The echo enclosed by a white rectangle is also shown as an inset in the upper right corner of the  figure. The inset shows the position of the IRS SL and LL slits.}}
\label{casa_echoes}
\end{figure}

One of the most unexpected discoveries made by {\it Spitzer} satellite has been 
the serendipitous detection of infrared (IR) light echoes in the vicinity of the 
Cas~A SNR. The possibility of detecting IR light echoes from ancient SNe in general, and Cas~A in particular, has been considered by \cite{van-den-bergh65} more than 4 decades ago.
Cas~A is the remnant of a supernova that exploded $\sim 320$ yr ago 
in the Perseus Arm of the Galaxy, 3.5 kpc away. The echoes were first (and best)
detected in 24~$\mu$m MIPS scans of Cas A (Fig. \ref{casa_echoes}). Subsequent observations revealed
that the echoes are also visible in all IRAC channels and ground--based $K$ band
observations. The echoes vary on time scales of $< 6$ months, and extend at 
least $1.5^{\circ}$ from the SNR \citep{hines04, krause05, kim08}. 
More recently, \cite{kim08} have used the 24~\mic\ maps of the echoes over a three year interval to probe the structure of the ISM around Cas~A, revealing its complex filamentary structure.

In this paper we combine early spectral observations with more recent data, that have revealed a wider extent of the echoes, to provide a detailed analysis of the nature of the radiative burst that gave rise to the echoes, and the density of the interstellar clouds giving rise to the echoes. 

We first present a brief review of the discovery of the echoes around Cas A, and derive the IR spectrum of select echoing clouds spanning a wide range of angular distances from the SNR (\S 2). We also provide a brief mathematical review of the geometry of an echo, and derive the distances of the echoing clouds from the center of the explosion. 
In \S3 we present the model input parameters: the burst delay time and its spectral characteristics, and the interstellar dust model used in the calculations. 
A stochastic heating model is used to calculate the thermal emission from the dust exposed to the burst of radiation, and in \S4 we describe the dependence of the calculated IR fluxes on the burst characteristics. The model fits are then used to derive the properties of the burst and the echoing clouds (\S5). The results of our paper are briefly summarized in \S6.

 \section{THE INFRARED ECHOES AROUND CAS A}

\subsection{The Discovery of the Echoes}
The initial MIPS 24 $\mu$m observations (a $0\fdg2\times0\fdg3$ scan map) 
of Cas A serendipitously revealed ``chains of IR knots'' 
extending up to $\sim 12'$ from the center of the SNR (Hines et al 2004).
The team followed up with ground-based K band (2.2~\mic) observations which revealed 
extremely high proper motions of the features. This motivated a second epoch
of MIPS 24~\mic\ observation that confirmed that much of the small scale 
structure outside of the Cas A SNR was changing on time scales of several 
months. A subset of the echoes seemed to represent a bipolar jet, nearly oriented in the plane of  the sky, leading  to the suggestion that at least these echoes may have been produced by a recent flaring of the Cas~A neutron star around 1952$\pm$2.5 A.D. \citep{krause05}, i.e. about 50~yr ago.  

Subsequent MIPS 24~\mic\ observations have been
repeated at roughly 6 month intervals, and have been expanded to include at first a 
$\sim1$ square degree field around the SNR, and later a $\sim9$ square degree 
field. Examination of these data from the Spitzer archive reveal that echoes 
are not restricted to the bipolar structure originally detected. They are found
at all position angles with respect to the SNR and extend at least $1\fdg8$
from the SNR (limited by the size of the map). The neutron star burst scenario was dismissed in passing by \cite{kim08}. Here we will provide additional physical evidence against such scenario.

Follow-up observations of the echoes also included low and high resolution IRS observations
of a bright echo in a filament located $0\fdg8$ from the center of the SNR.
Additional low resolution spectra were collected for outlying regions in 
the vicinity of the SNR. Two of the closest regions to the SNR, located within the NE jet of ejecta and 
the SW counter jet, have emission lines and the 22~\mic\ silicate 
feature that are characteristic of ejecta material \citep{ennis06,rho08}. Spectra of other regions exhibited a flat continuum and coincide with highly variable features, and are therefore assumed to be representative echo spectra. \footnote{Archival MIPS images were obtained from Spitzer programs 
PID = 718, 30571 -- G. Rieke, and PID = 231, 233, 20381 -- O. Krause.}

\begin{figure}[htbp]
\hspace{-0.3in}
	\includegraphics[width=7.0in]{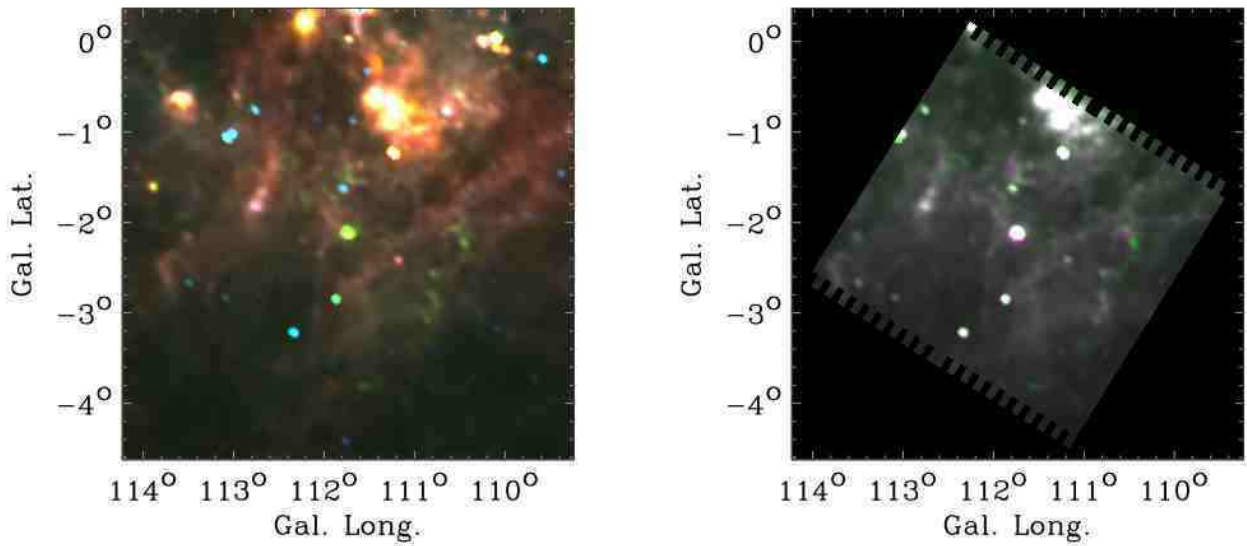}
  \caption{{\footnotesize
 {\bf Left panel:} This false color image displays the region around Cas A as observed by \iras\
in 1983. Reprocessed \iras\ data (IRIS) at 60, 25, and 12 \mic\ are
depicted as red, green, and blue respectively. The extended structures
with excess 25 \mic\ emission (green in color) are now know to be light
echoes as they have significantly changed in \spitz\ MIPS images in 2006.
Cas~A is the bright green source at the center of the image. {\bf Right panel:} This image shows IRAS 25 \mic\ observations from 1983 in green, and
\spitz\ MIPS 24 \mic\ observations from 2006 in magenta. The MIPS
data have been smoothed to approximate the IRAS spatial resolution.
Cas A is the bright compact source at the center of the map. The light echoes
are visible as extended structures, with magenta or green tints depending on
when they were present. Nonvariable structure appears a uniform gray shade.
Most of the 1983 (green) echoes seen here are also evident via their
mid-IR colors as revealed the the previous figure.
}}
\label{casa_iris}
\end{figure}

The discovery of the echoes with the \spitz\ led us to examine if the echoes could also be identified (in hindsight) in the map of the Cas~A region that was obtained in 1983 with the {\it Infrared Astronomical Satellite} (\iras). Figure \ref{casa_iris} (left panel) shows a false color image at 12, 25, and 60~\mic\ of the reprocesses \iras\ map of Cas~A and its environment. The emission is dominated by cool interstellar dust grains, peaking  at wavelengths of $\sim 100-140$~\mic\ (red regions in the image). 
However, there are several regions with excess 25~\mic\ emission (including Cas~A itself) from hot dust, which thus appear green. Comparison to the 24~\mic\ MIPS data (right panel), smoothed to \iras\ resolution, 
clearly shows that the 25~\mic\ excess features are echoes which have moved between 1983 and 2003. The fact that  many echoes are in similar locations indicates that many of the clouds producing the echoes are structures that extend for at least 20~lyr along  the line of sight (cf. Kim et al. 2008). \iras\ and \spitz\ have observed different portions of these large-scale structures.

\subsection{The IR Spectrum of Select Echoes}

\begin{figure}[htbp]
  \centering
  \includegraphics[width=5.0in]{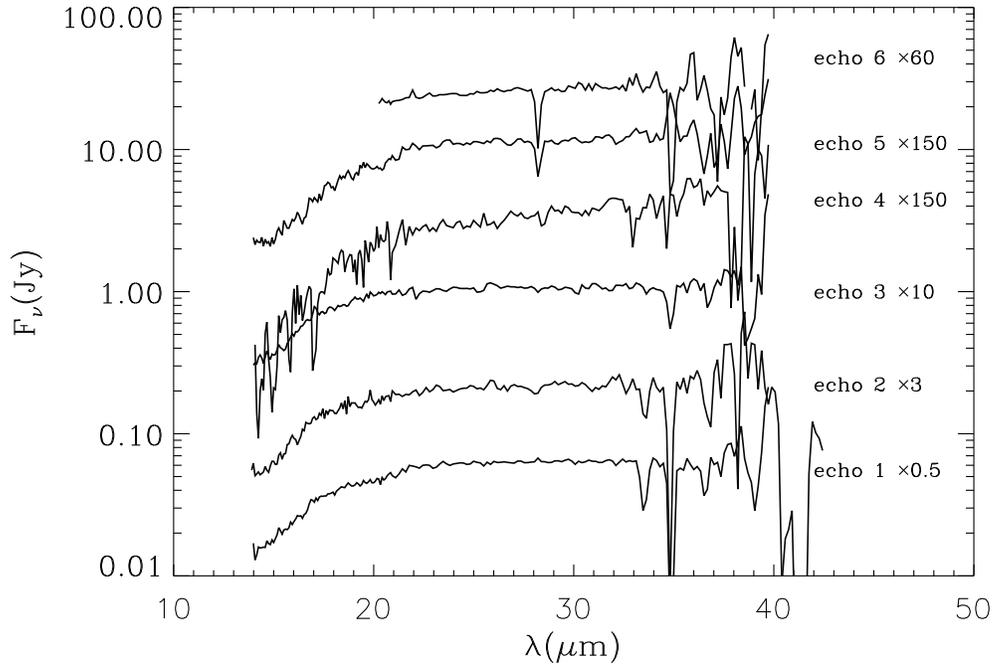}
  \caption{{\footnotesize
The observed spectrum of the first six echoes listed in Table~1. 
 }}
\label{echo_data}
\end{figure}

For the purpose of this work we have concentrated on the analysis of six IR knots that were identified as echoes and have been the target of IRS observations. The selected echoes, listed in Table 1 and shown in Figure \ref{casa_echoes}, sample a range of angular distances from the nominal center of the explosion.
Also included in the table is a seventh 24~\mic\ echo, without a spectrum, as a representative of the most distant echoing cloud. The SL+LL IRS spectra of the echoes were collected from the Spitzer archive.\footnote{Archival IRS spectra were obtained from Spitzer programs PID = 3310 -- L. Rudnick, and PID = 20381 -- O. Krause.}
Because there are often several sources or extended emission within the slits,
we have reprocessed the basic calibrated data (BCD) using the SSC tools
IRS\_CLEANMASK and SPICE. The off-target subslit positions were used to subtract
background emission for the corresponding on-target position. In a few 
cases where the off-target subslit position crossed the SNR, a clean subslit 
position from an alternate target was used. 
Both SL and LL data were processed, although in most cases the SL integration times were too short to detect the echoes, which are visible in all IRAC bands. Therefore the analysis 
presented here relies only on the LL spectra.
Figure \ref{echo_data} presents the $\sim 14 - 40$~\mic\ spectra of the echoes listed in Table~1. 

All echoes within a 1000$''$ of Cas~A exhibit a steep rise in their spectrum between $\sim 14 - 20$~\mic, not seen in the IR spectra of dust in the diffuse ISM that is heated by the general interstellar radiation field \citep{draine07}. This spectral characteristic provides important clues on the nature of the emitting dust, strongly constraining the viability of different echo models. Echo~6 is brighter than the echoes closer to the SNR, and is part of a large extended structure that can be seen in the older IRAS data.
This region appears to contain a non-variable emission component which would simply be the same cloud(s)
illuminated by the diffuse interstellar radiation field. We have therefore subtracted a scaled ISM
spectrum from the Echo~6 spectrum such that its 20 -- 35 $\mu$m slope is within the range of the slopes of the other echo spectra.
This non-variable component accounted for 20 -- 50\% of the 20 -- 35 $\mu$m emission.

 \subsection{The Location of the Echoes}

\begin{figure}[ht]
  \centering
    \includegraphics[width=4.0in]{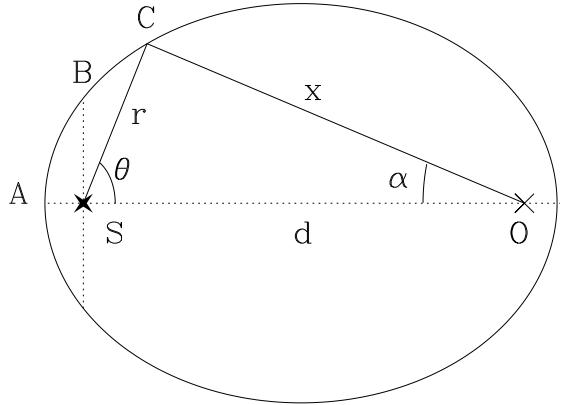}
  \caption{{\footnotesize
 The geometry of an echo. The locus of all points with equal delay time is described by an ellipsoid with the source "S" and the observer "O" at the focal  points. The lengths $\overline{AS} = ct/2$, and $\overline{BS} = ct$, where $t$ is the difference between the observing and the emission time of a given photon. The point "C" is an arbitrary point on the ellipse sustending an angular distance $\alpha$ from the source, located at distances $r$ and $x$ from the source and observer, respectively. The length $\overline{OS} = d$, the distance of the source from the observer.}}
\label{ellipse}
\end{figure}

We assume that the echoes are the reradiated thermal emission from dust heated by the UV-visual output from the SN, located at point $S$ (see Figure \ref{ellipse}). 
 The emission from dust located at point $C$  will be detected by an observer at point $O$ with a delay time $t$ given by $ct=r+x -d$, where $r$ and $x$ are, respectively, the distances of $C$ from the source and the observer, and $d= \overline{OS}$ is the distance of the observer to the source. The locus of all points with equal delay time is an ellipsoid with the source and the observer at the focal points. 

Using the law of cosines: $r^2=x^2+d^2-2xd\cosae$, we can express the distance $r$ in terms of the delay time $t$ and source distance $d$ as:
\begin{equation}
\label{dist}
r = {d (d+ct)(1-\cos(\alpha))+(ct)^2/2 \over d (1-\cos(\alpha))+ct}
\end{equation}
where $\alpha$ is the angular distance of the echoing dust from the source.

\begin{figure}[htbp]
  \centering
  \includegraphics[height=2.5in]{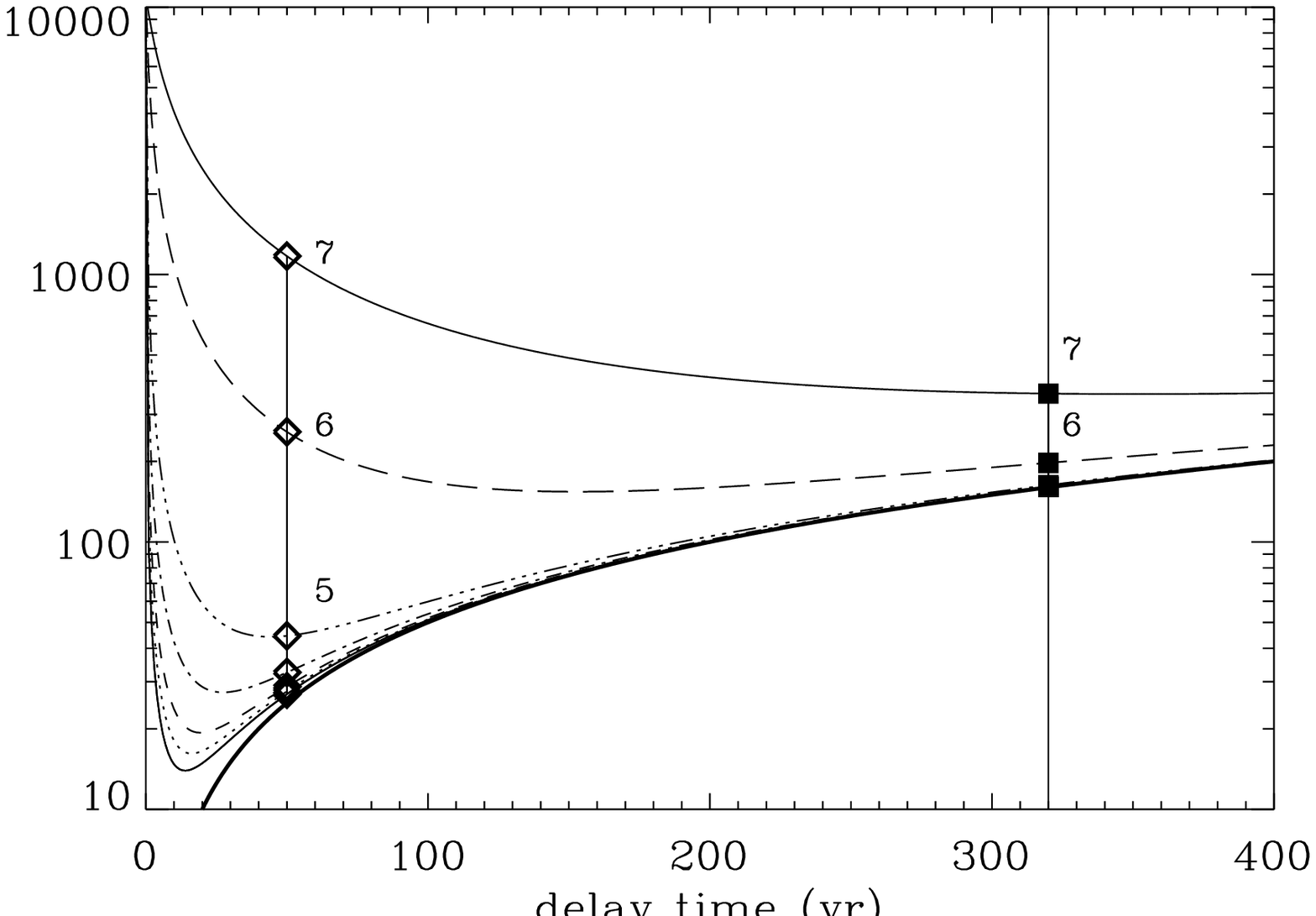} \\
  \vspace{0.1in}
    \includegraphics[height=2.5in]{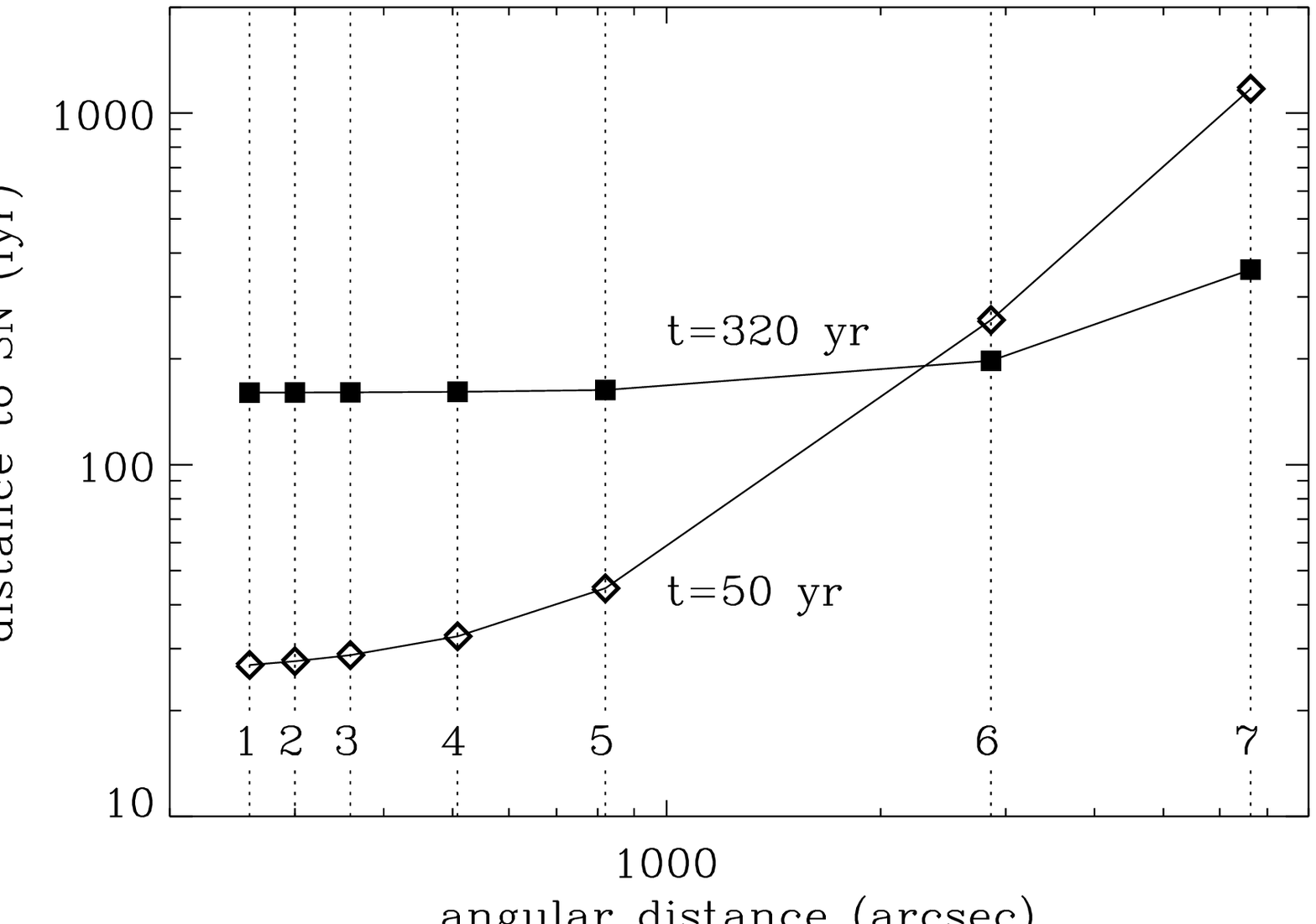} \\
    \vspace{-1.5in}
      \includegraphics[width=5.0in]{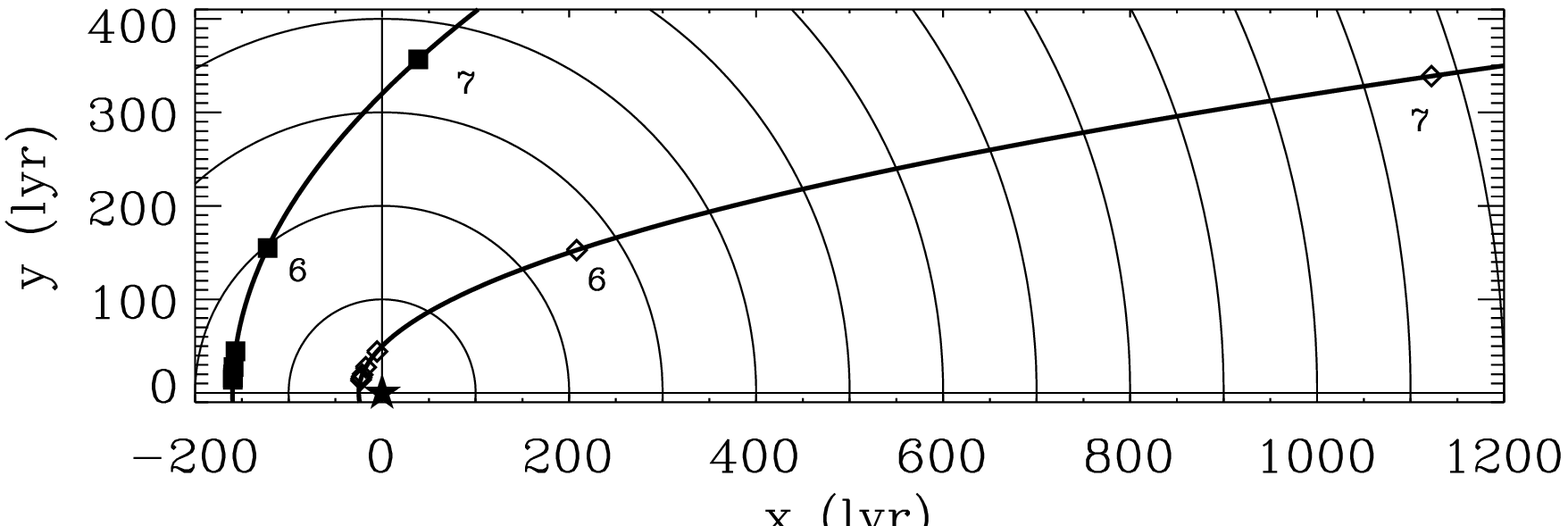}
  \caption{{\footnotesize
Three different depictions of the interrelations between the delay time, and the angular and physical separations of the echoing clouds listed in Table 1 from the source. The relation between the different quantities is given by  eq. (\ref{dist}) for $d = 3.4$~kpc, the distance to Cas~A. {\bf Top:} Echo distance to the source  versus the delay time for different angular separations. {\bf Middle:} echo distances to the source as a function of angular distance for delay times of 50 and 320~yr. {\bf Bottom:} Physical location of the echoes with respect to the source, as projected on a plane that includes the source, located at $(0, 0)$,  and the observer (off scale), located at $(d, 0)$. All figures are described in more detail in \S2.3 of the text.}}
\label{dist_del_ang}
\end{figure}

Figures  \ref{dist_del_ang}a-c are three different depictions of the interrelations between the delay time, and the angular and physical separations of the echoing clouds from the source assuming  a source at the distance of Cas~A, corresponding to $d = 3.4$~kpc \citep{reed95}. Figure~\ref{dist_del_ang}a shows the echo distance from the source, $r$, as a function of the delay time $t$ for the separation angles $\alpha$ corresponding to the echoes listed in Table 1. 
All distances are bound by the lower limit $ct/2$ which is the minimum distance corresponding to dust located at point $A$ immediately behind the source (see Figure \ref{ellipse}).  The vertical  lines reprsents the maximum delay time, corresponding to $t = 320$~yr, the age of Cas~A, and a delay time of 50~yr. Figure \ref{dist_del_ang}b depicts the echo distance from the source as a function of angular separation. The vertical dashed lines correspond to the angular separations of the echoes listed in Table 1. The filled squares and open diamonds correspond to distances calculated for delay times of $ct = 320$ and  50~yr, respectively. Figure \ref{dist_del_ang}c depicts the physical location of the echoes with respect to the source, as projected on a plane that includes the source, located at the origin $(x, y) = (0, 0)$,  and the observer (off scale), located at $(x, y) = (d, 0)$.  The vertical line going through the origin represents the plane of the sky going through the source. The two ellipses shown in the figure are the loci of all points with equal delay times. The filled squares and open diamonds represent the echoes listed in Table 1 assuming delay times of 320 and 50~yrs, respectively. Echo 7 with  the largest separation angle from the source is almost on the plane of the sky when $t = 320$~yr, but at a large distance in front of the source when $t = 50$~yr. 

The figures illustrate that when delay times are short, a given range of angular distances will correspond  to a wide range of echo-source distances, whereas long delay times will translate ino a significantly narrower range of echo-source distances for the same range of angular distances. When $t = 50$~yr, the range of distances, $r$,  spanned by the echoes in Table~1 is between 27 and 1172 lyr. This range is narrowed to only 160 and 359~lyr when $t = 320$~yr (see Table 2).    

\section{MODEL INPUT PARAMETERS}

\subsection{The Echo Delay  Time}

The most obvious explanation for the echoes is that they were generated by the radiative energy released following the core collapse of the progenitor star of the Cas~A SN, which occurred about 320~yr ago. Since \cite{krause05} suggested that a subset of the echoes were produced by a recent  flaring of the neutron star that happened around 1950, we will also examine if such outburst could have produced the other echoes as well. 

For each of the two scenarios ($t$ = 50 or 320~yr) we determine the incident flux required to produce each echo spectrum.  Since the two scenarios place the echoing dust clouds at different distances, these fluxes will translate into different burst luminosities. To check the viability of any physical scenario we require that all echoes are generated by a single burst with a well-defined intrinsic luminosity. 

It is possible to construct scenarios involving multiple bursts from a neutron star over the past 320 years. However, such scenarios have too many free parameters to be usefully constrained by the observations. Furthermore, the construction of such scenarios seems unnecessary given the consistent results that are ultimately obtained with the 320~yr old SN echo scenario. 

\subsection{The Burst Spectrum and Duration}    
Because neither the Cas~A SN nor any subsequent outburst of the neutron star has been observed, we do not know the spectrum of the radiation that gave rise to the echoes. We will therefore explore three possible "burst" spectra: (1) an EUV spectrum characterized by a temperature of $T_b = 5\times10^5$~K; (2) a UV spectrum characterized by a temperature of $T_b = 5\times10^4$~K; and (3) an optical spectrum characterized by a temperature of $T_b = 6000$~K. 

The radiative burst giving rise to the IR echo could be     
generated by the breakout of the SN shock wave through the stellar surface, in which case the burst is powered by the kinetic energy of the explosion. 
A flash of EUV-UV shock breakout radiation has been predicted by hydrodynamical simulations of SN explosions \citep{klein78, falk78}, however, because it lasts for only a few days, it has only recently been directly observed in a SN. Its existence was first indirectly inferred in SN1987A from the detection of narrow UV and optical emission lines from the SN \citep{fransson89a, sonneborn97}. The lines arise from an equatorial ring, located about 0.7 lyr from the SN and produced by mass loss from the progenitor star that was flash-ionized by the EUV burst. More recently, \cite{soderberg08}, using the {\it Swift} X-ray Telescope, detected an X-ray burst from SN~2008D which they attributed to the shock breakout.

Alternatively, the IR echo could be powered by the energy released in the decay of radioactive elements that have formed in the SN explosion. In addition to the optical emission, a significant UV component has been observed in the light curves of all major types SNe with the {\it Swift} UV-optical Telescope \citep{brown08}.  
  
The IR spectrum of the echo depends on the spectral shape and intensity of the flux incident on the dust. The luminosities of the burst are free parameters that will be derived from model fits to the IR echoes. The duration of the burst is constrained by energy considerations and only affects the thickness and overall brightness of the expanding light echo. The burst duration is a necessary parameter in converting  the derived column density into a volume density of the radiating dust.

\subsection{Interstellar Dust Model}
We used the  BARE-GR-S model of \cite{zubko04}  to characterize the echoing interstellar dust. The model consists of a population of bare silicate and graphite grains and PAH molecules (see Table 5 in their paper). In this model, the dust-to-gas mass ratio is 0.0062, and the abundance of its various dust constituents is consistent with solar abundances constraints. The size distribution of the different dust species are depicted in Figure~19 of their paper, and the parameters of their functional forms are listed in their Table~7.     

\section{MODELING THE ECHO SPECTRA}
 
\subsection{The IR Emission from Stochastically Heated Dust}
The observed IR flux, $F_{\nu}(\lambda)$, from an extended source of radiating dust is given by:
\begin{equation}
\label{fnu}
F_{\nu}(\lambda) = \left({\Omega\over\pi}\right)\, {\cal M}_d\,  \left[\, \int_{a_1}^{a_2}\, f(a)\, \kappa(\nu, a)\, {\cal G}_{\nu}(\lambda, a)\, da \right]
\end{equation}
where $\Omega$ is the angular size of the dust cloud, ${\cal M}_d$ is the mass column density of the radiating dust, $f(a)$ is the grain size distribution normalized to unity over the \{$a_1, a_2$\} size interval, and $\kappa(\nu, a)$ is the mass  absorption coefficient of the dust. Implicit in eq. (\ref{fnu}) is a sum over dust composition. The term $\kappa(\nu, a)\, {\cal G}_{\nu}(\lambda, a)$ in the equation is the spectral response of the dust to an incident radiation field, with ${\cal G}_{\nu}$ given by an integral over dust temperatures $T_d$:
\begin{equation}
\label{ gnu}
{\cal G}_{\nu}(\lambda, a) \equiv \int_{T_d}\, P(T_d)\,  \pi B_{\nu}(T_d, \lambda)\, dT_d
\end{equation}
where $P(T_d)$ (K$^{-1}$) is the temperature probability distribution of the stochastically heated dust, and  $B_{\nu}(T_d,\lambda)$ is the Planck function. For dust radiating at the equilibrium temprature $T_d^{eq}$, $P(T_d)$ collapses to the delta-function $\delta(T_d^{eq})$, and ${\cal G}_{\nu}(\lambda, a) =  \pi B_{\nu}(T_d^{eq}, a)$

\subsection{Echo Spectra For Different Burst Models}
To model the IR emission from the echoing clouds we will expose a population of interstellar dust to a burst of radiation emanating from the SN. The burst spectrum will be characterized by a blackbody at a fixed temperature $T_b$. A significant fraction of its soft X-rays and EUV photons will be absorbed in the intervening medium to the clouds. The optical depth, $\tau(\nu)$, will be parametrized by  the intervening  ISM column density, $N_H$. The flux incident on the dust  grains is then given by:
\begin{equation}
\label{flux}
F_{\nu}^b(\lambda)  =  \xi\, F_0\, \left[{\pi B_{\nu}(T_b, \lambda)\over \sigma T_b^4}\right]\, e^{-\tau(\lambda)} \qquad ,
\end{equation}
where $F_0$ is a fixed flux, calculated for a nominal luminosity $L_0= 1\times 10^{12}$~\lsun\ and a nominal distance of $r_0 = 160$~lyr between the radiating dust and the source, and $\xi$ is a flux scaling parameter determined by the fit to the observed spectrum of the IR echo. Numerically, $F_0$ is given by:
\begin{equation}
\label{ }
F_0  \equiv  {L_0\over 4 \pi r_0^2} = 1.33\times10^4\ {\rm erg\ s}^{-1}\, {\rm cm}^{-2}\qquad .
\end{equation}

Since the H-column density to the echoing clouds is unknown, we will calculate the burst's attenuation for three different H-column densities given by $N_H = 1.5\times10^{18}, 1.5\times10^{19}$, and $1.5\times10^{20}$~cm$^{-2}$. For a typical echo  distance of 160~lyr these column densities correspond to average ISM densities of $n_H = 0.01,\ 0.1$ and 1.0~\cc, respectively. 

Our burst model calculations were performed for seven different burst spectra characterized by different burst temperatures and intervening H-column densities. 
Burst models are designated by their spectra: EUV for $T_b = 5\times 10^5$~K; UV for $T_b = 5\times 10^4$~K; and \olc\ for $T_b = 6000$~K, and by a suffix of 18, 19, or 20, corresponding to the assumed SN-echo column density.
For example, model \euvb\ corresponds to a burst with a temperature of $5\times 10^5$~K and a column density of $1.5\times10^{19}$~cm$^{-2}$ to the echoing clouds. Model \uvc\ corresponds to a burst with a temperature of $5\times 10^4$~K and an ISM column density of $1.5\times10^{20}$~cm$^{-2}$. 
We neglected any attenuation for the \olc\ burst. This model lacks any significant ionizing radiation, and therefore the extinction was negligible even for the largest column density. 

We calculated echo spectra for each burst spectrum for a grid of fluxes, characterized by different values of $\xi$ ranging from $10^{-3}$ to 10. For example, we calculated 16 IR spectra for burst model EUV18 by  varying $\xi$ from 0.001 to 1.0 with 14 intermediate values. The only free model parameter is the normalization factor, which is equal to the dust mass column density, ${\cal M}_d$, of the model spectrum that produced the lowest value of \chisq.  

\begin{figure}[ht]
  \centering
    \includegraphics[width=4.0in]{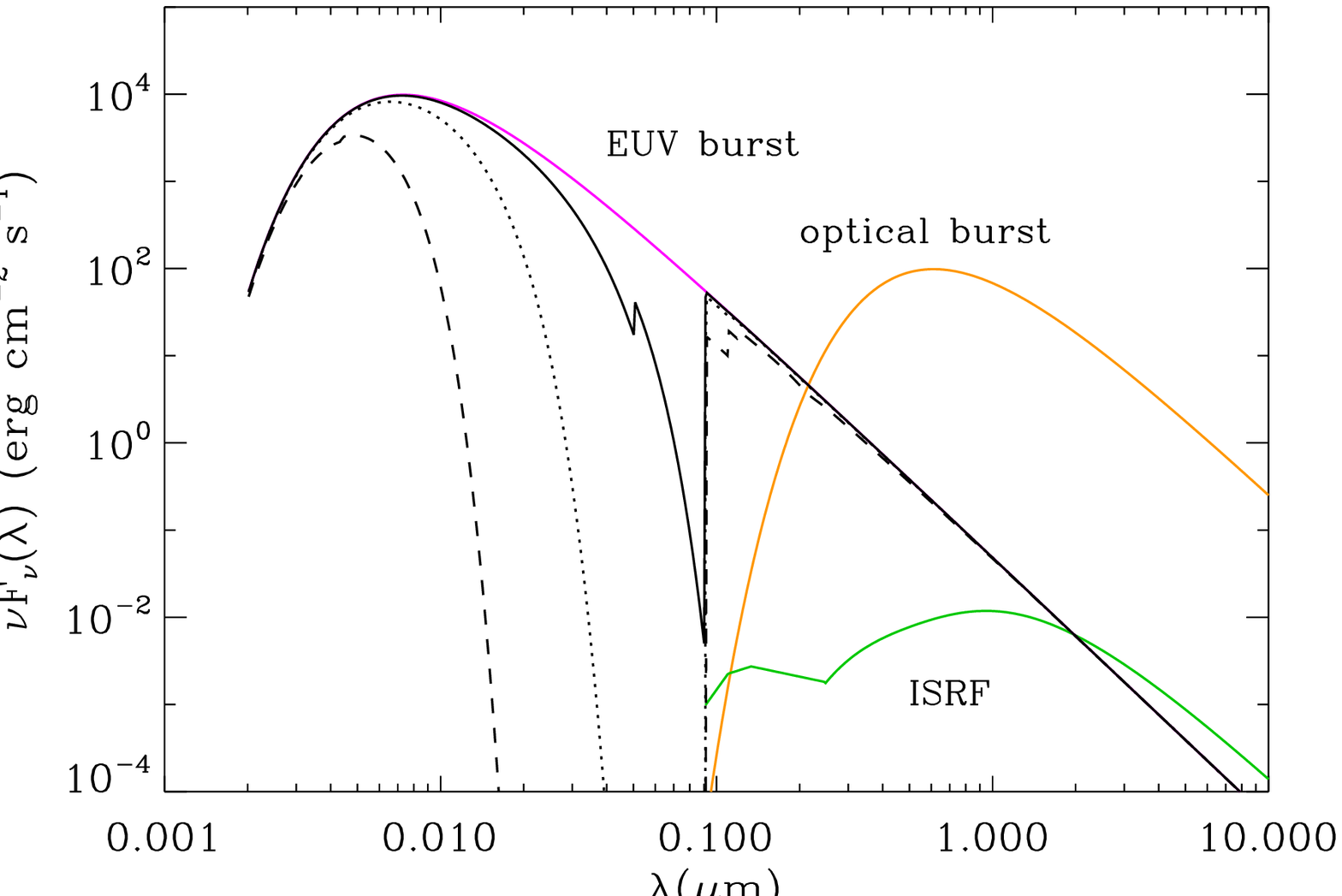}\\
      \includegraphics[width=4.0in]{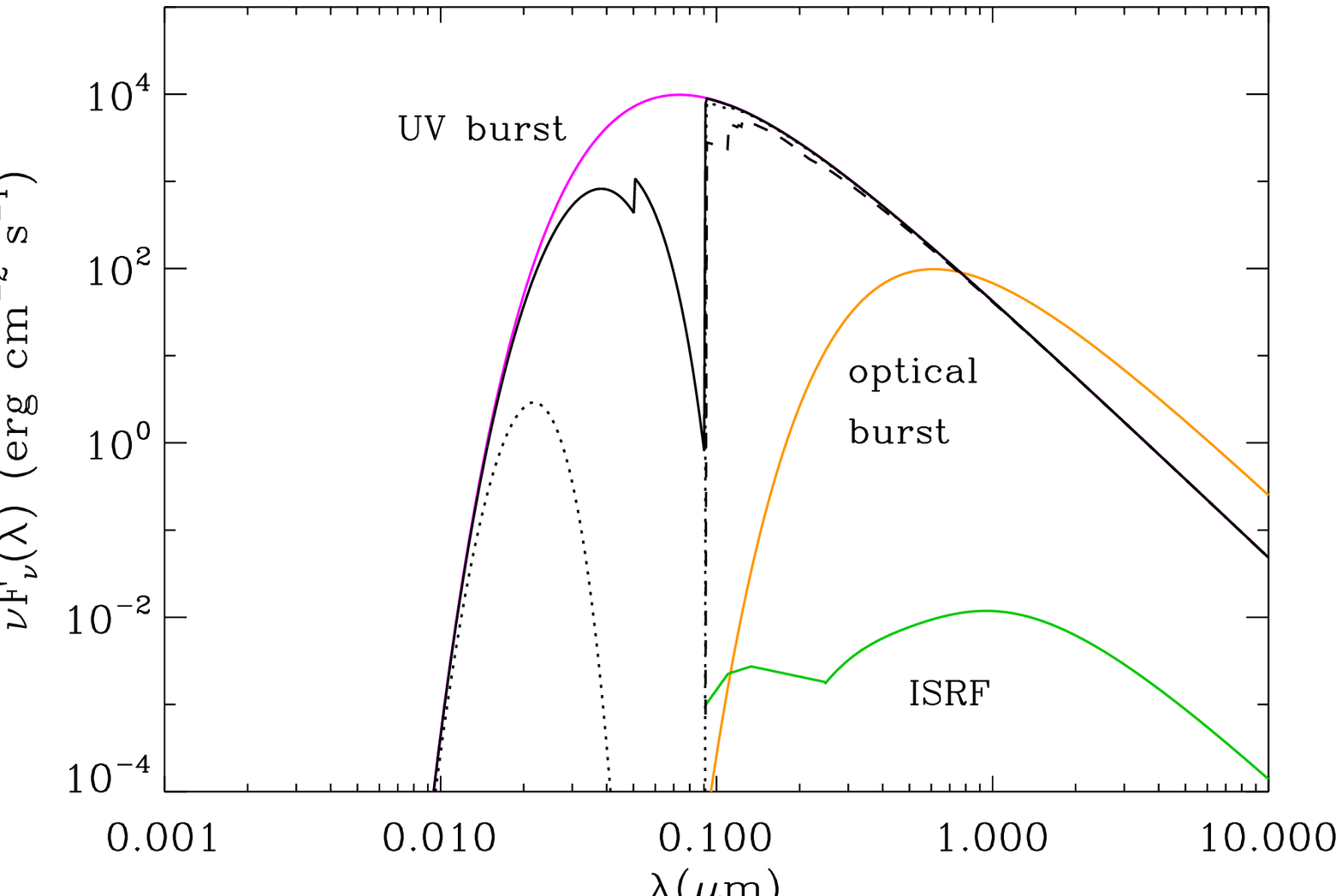}
  \caption{{\footnotesize
 The flux of the different bursts considered in this paper (\S4.2): the EUV burst characterized by a blackbody temperature of $T_b=5\times10^5$~K (top panel), and the UV burst with $T_b=5\times10^4$~K (bottom panel). Also shown in the figure is the flux from the optical burst with $T_b=6000$~K. For comparison, we also included the flux of the ISRF.
The intrinsic luminosities of the EUV and UV burst were taken to be $10^{12}$~\lsun. The luminosity of the optical burst was taken to be $10^{10}$~\lsun. All fluxes were calculated for a distance of 160~lyr 
The unattenuated burst is shown as a violet line, and the solid, dotted, and dashed line represent the flux attenuated en route to the echoing cloud by the ISM with column densities of $1.5\times10^{18}$, $1.5\times10^{19}$, and $1.5\times10^{20}$~cm$^{-2}$, respectively. Detailed burst parameters are described in the text.
 }}
\label{burst_spec}
\end{figure}

Figure \ref{burst_spec} compares the fluxes from the different burst scenarios outlined above as they are incident on an echoing cloud at a distance of 160~lyr from the SN. For illustrative purposes, the fluxes from the EUV and UV bursts were calculated for values of $\xi = 1.0$, and the flux from the optical burst was calculated for $\xi = 0.01$, reflecting its expected lower luminosity. For sake of comparison, the figure also includes the flux of the general interstellar radiation field [ISRF; \cite{mathis83}].    

\begin{figure}
  \centering
  \vspace{-0.3in}
   \includegraphics[width=3.5in]{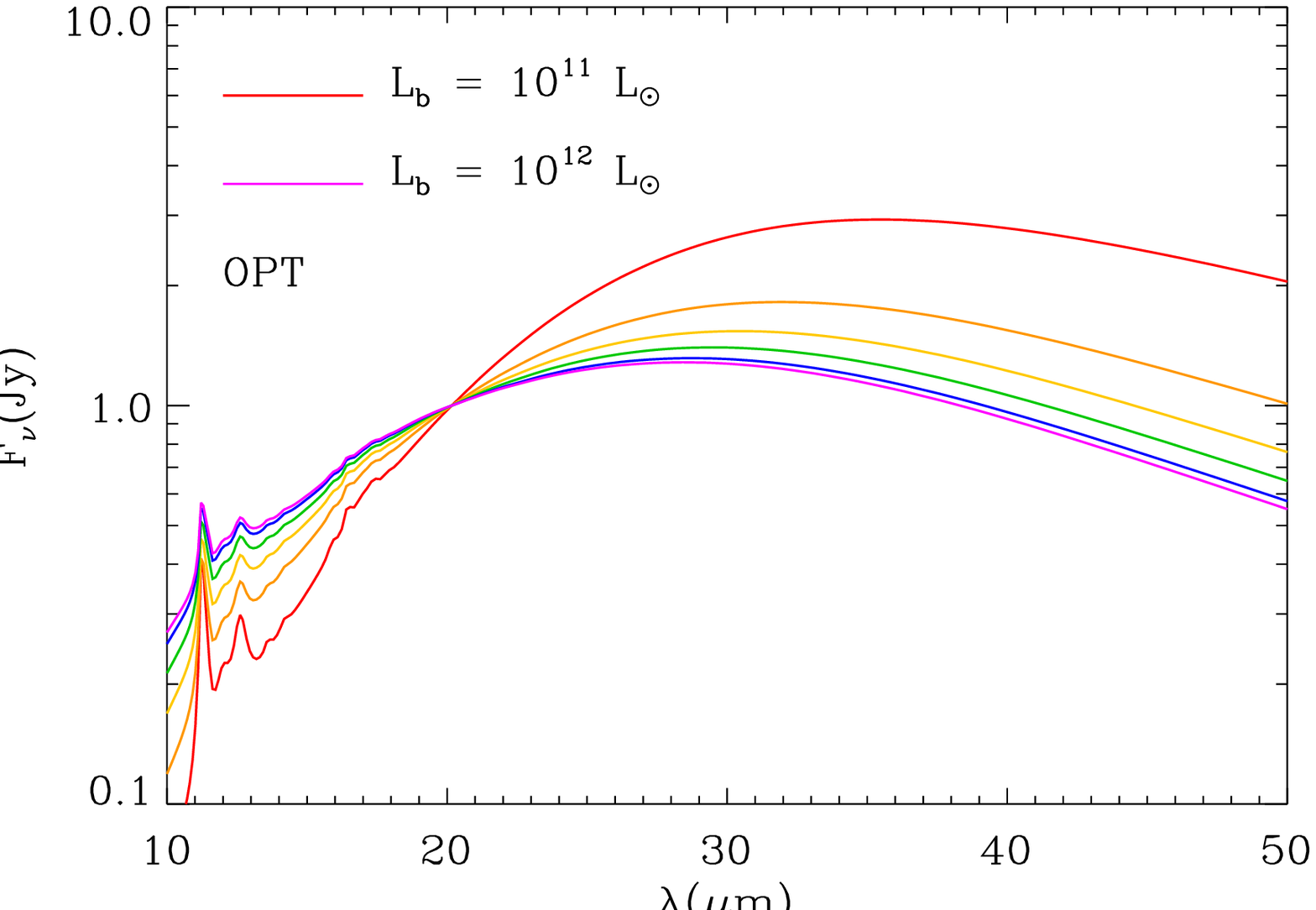}
  \includegraphics[width=3.5in]{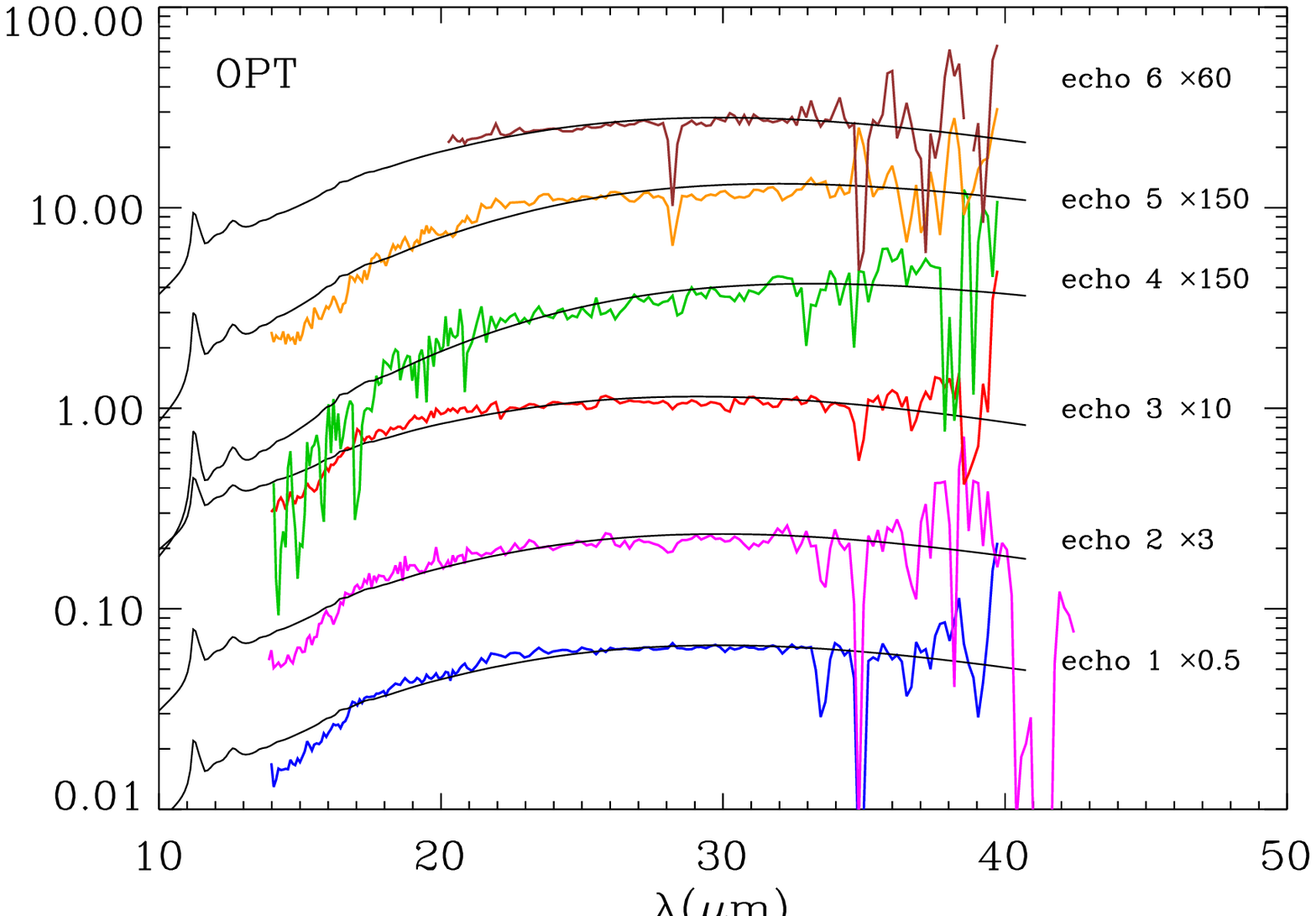} \\
   \includegraphics[width=3.5in]{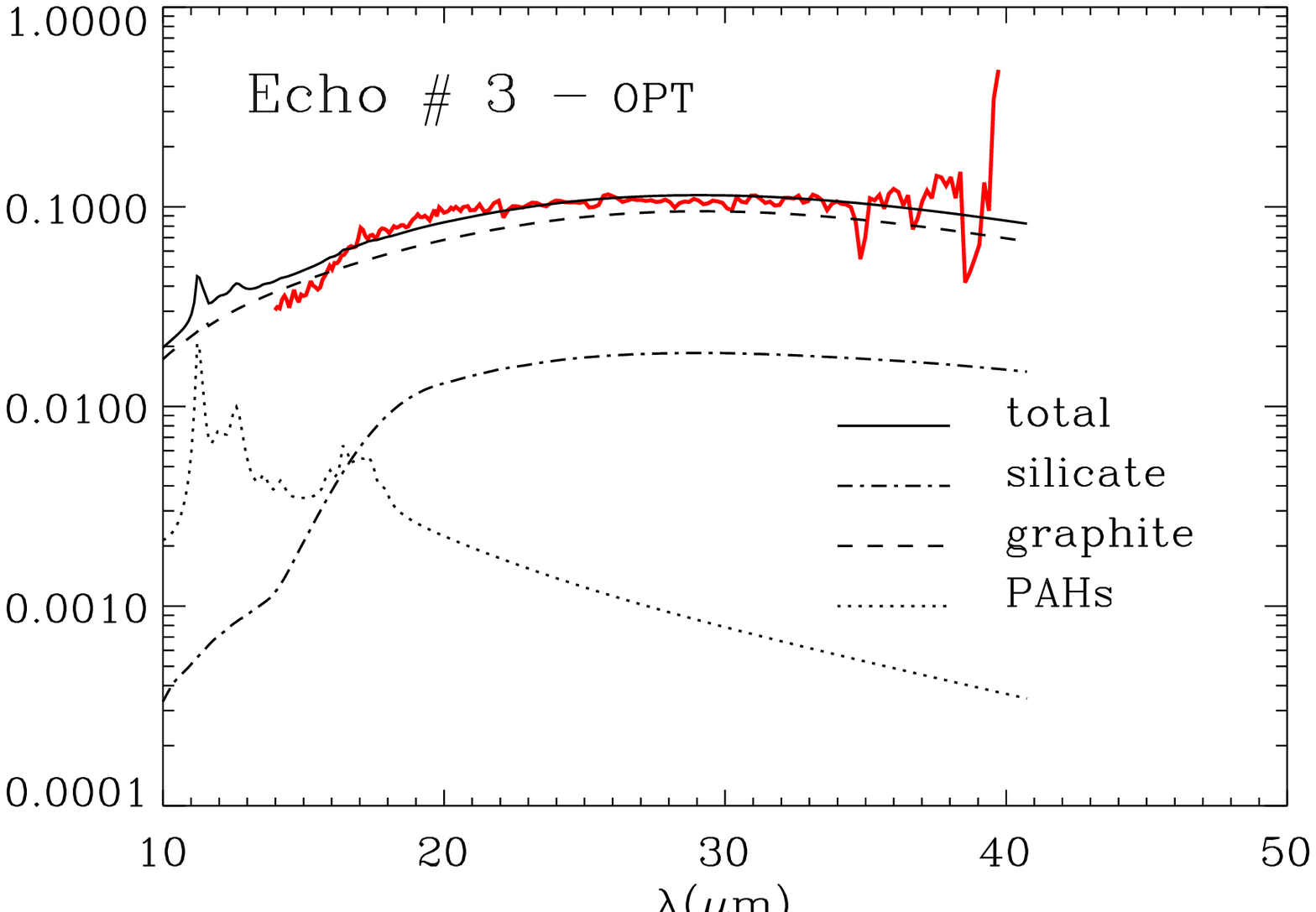}  \\
  
  \caption{{\footnotesize
Model fits for the optical burst (\olc). {\bf Top row:} Calculated IR spectra of an echo exposed to different fluxes $F_{\nu}^*$ characterized the product $\xi\, L_0$ and calculated for $r_0 = 160$~lyr (see eq. \ref{flux}) with $\xi$ = 0.1, 0.3, 0.5, 0.7, 0.9, and 1.0. All fluxes were calculated for $r_0 = 160$~lyr (see eq. \ref{flux}) and are normalized to unity at $\lambda = 20$~\mic. {\bf Middle row:} Best fit models to the IR echoes. {\bf Bottom row:} Decomposition of the best fitting IR spectrum of Echo~3 into its emission components from silicate and graphite grains and PAHs. The figure illustrates the failure of the \olc\ burst to fit the echo spectra. A detailed discussion of the figure is in the text.}}
\label{olc}
\end{figure}

\begin{figure}
  \centering
  \vspace{-0.3in}
  \includegraphics[width=2.1in]{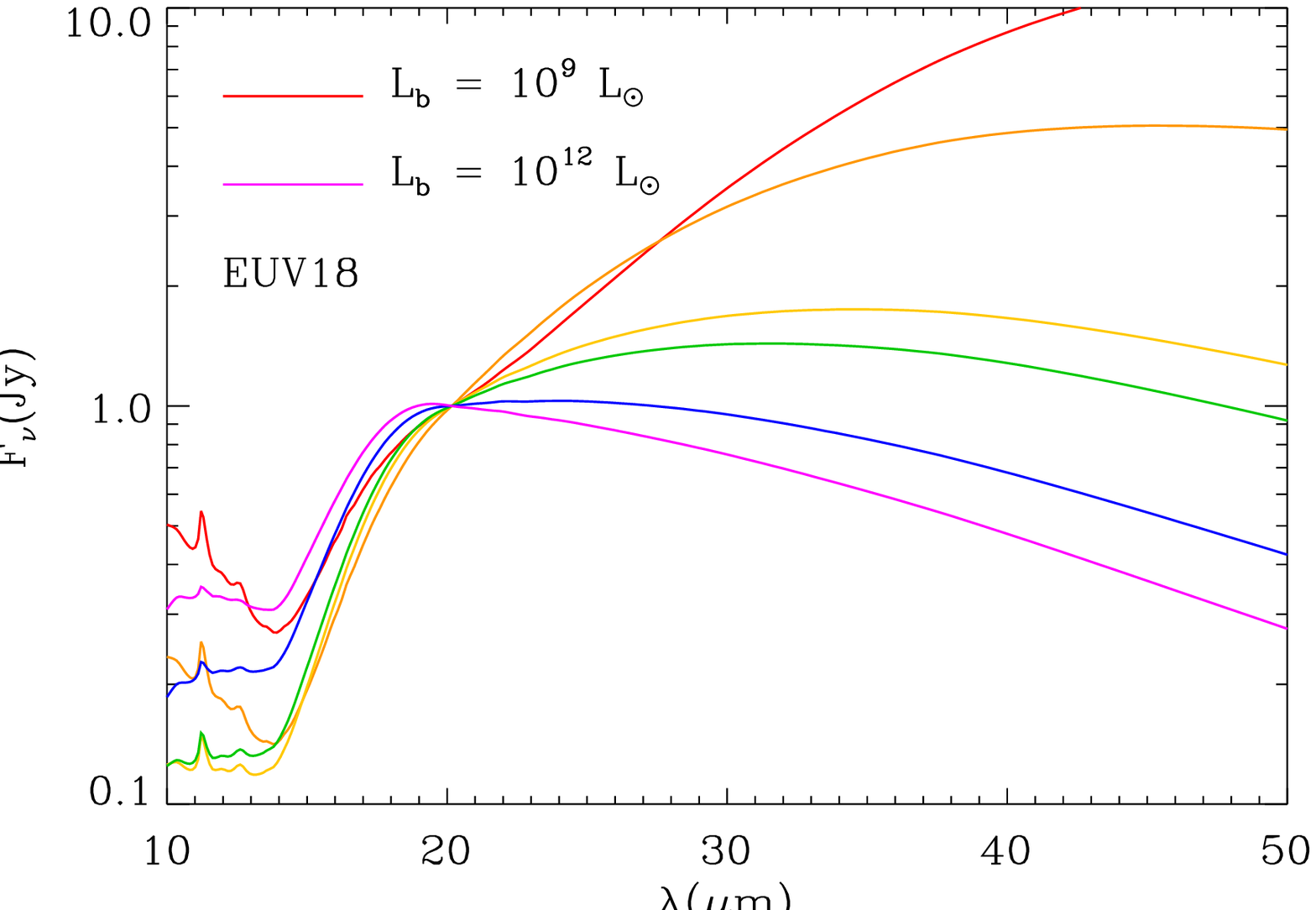}  
   \includegraphics[width=2.1in]{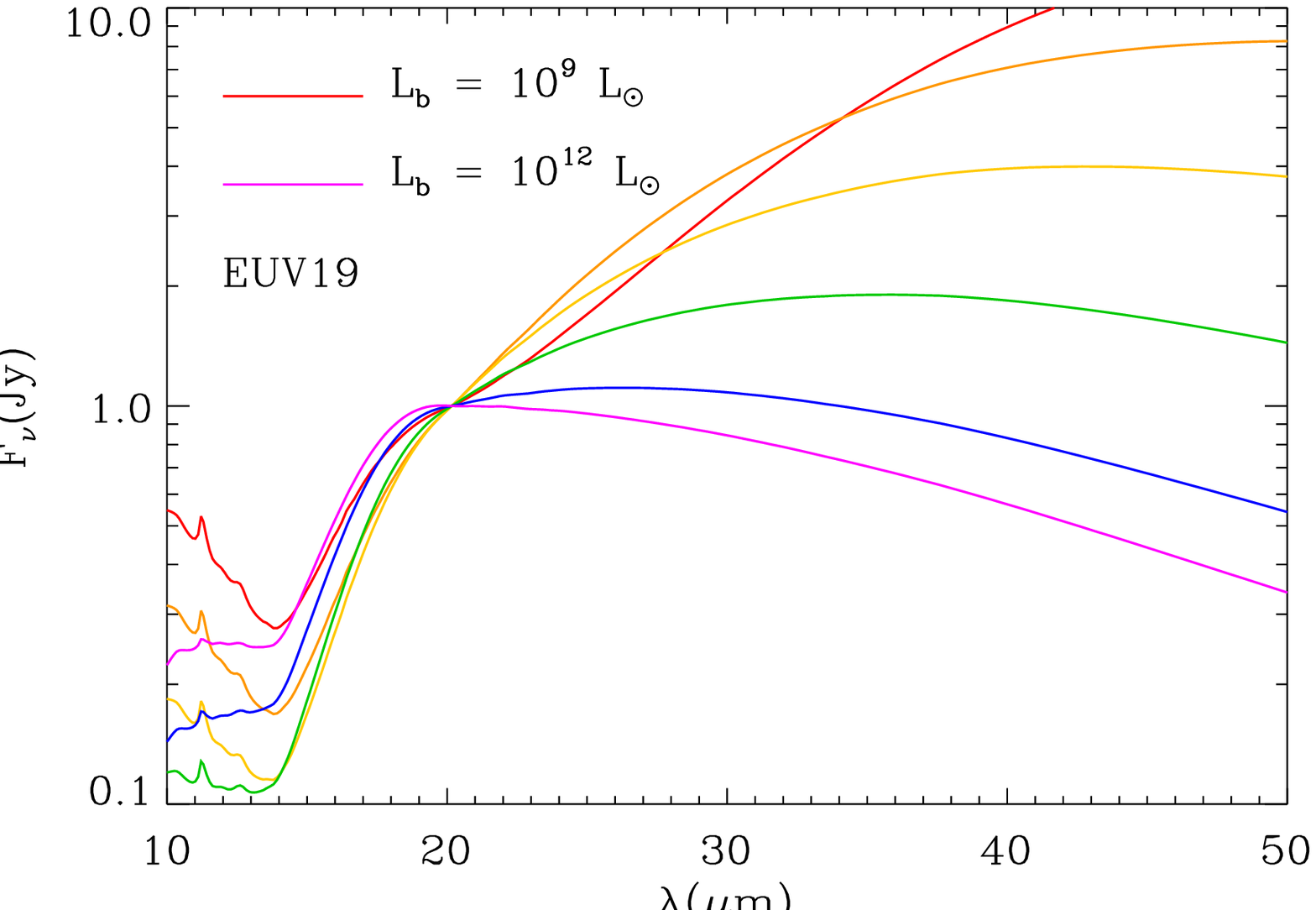} 
    \includegraphics[width=2.1in]{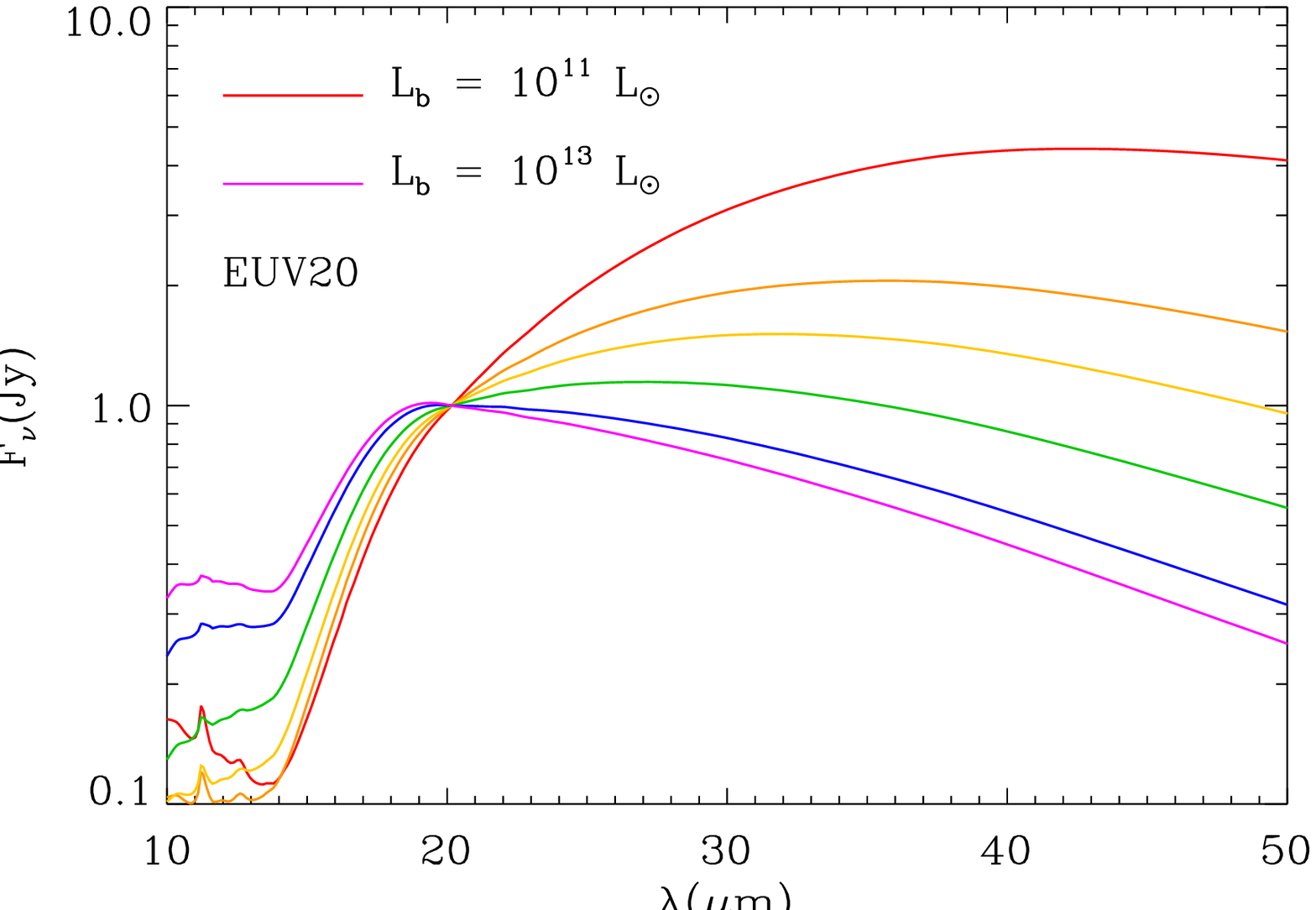} \\
  \vspace{0.1in}
    \includegraphics[width=2.1in]{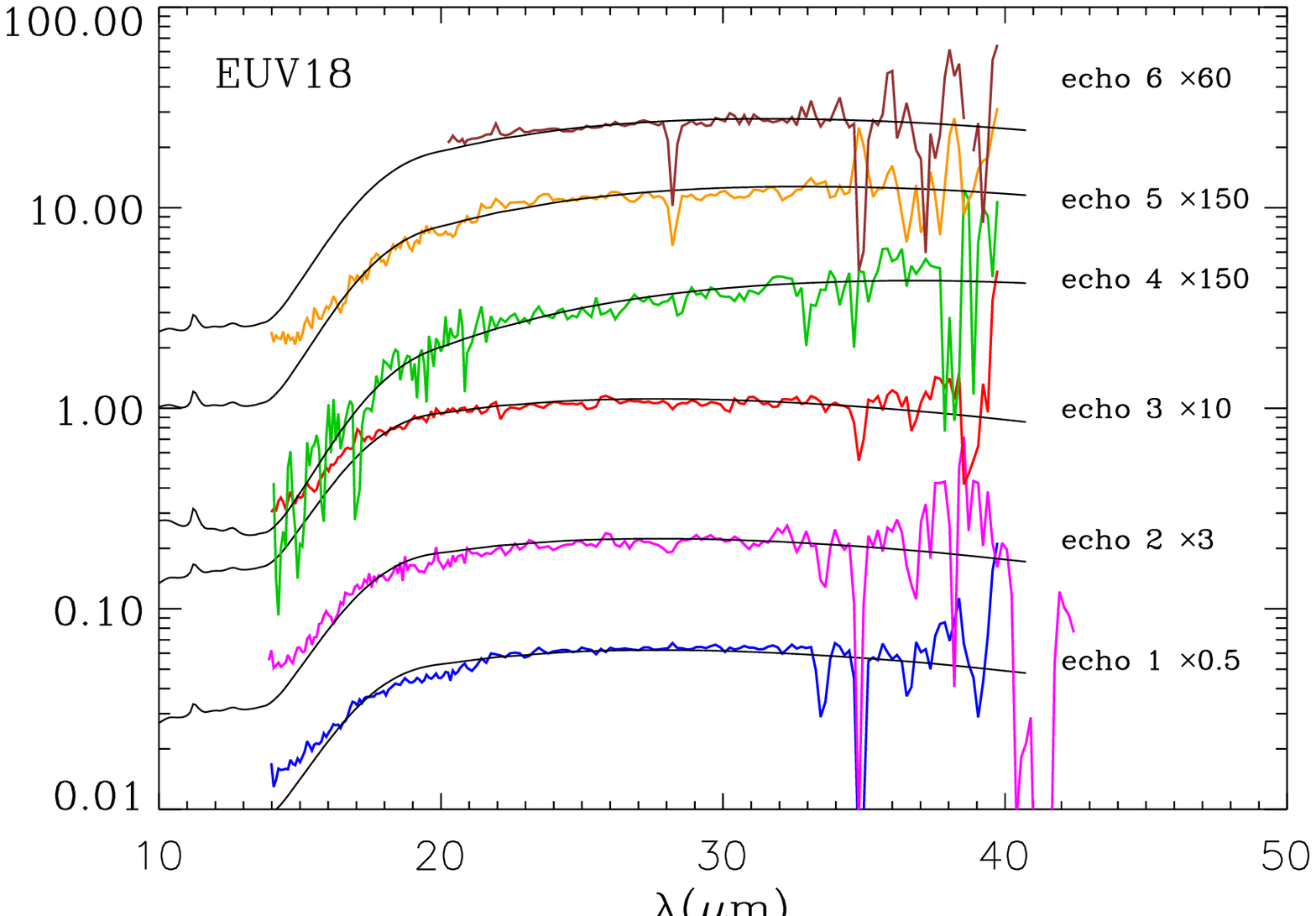}
       \includegraphics[width=2.1in]{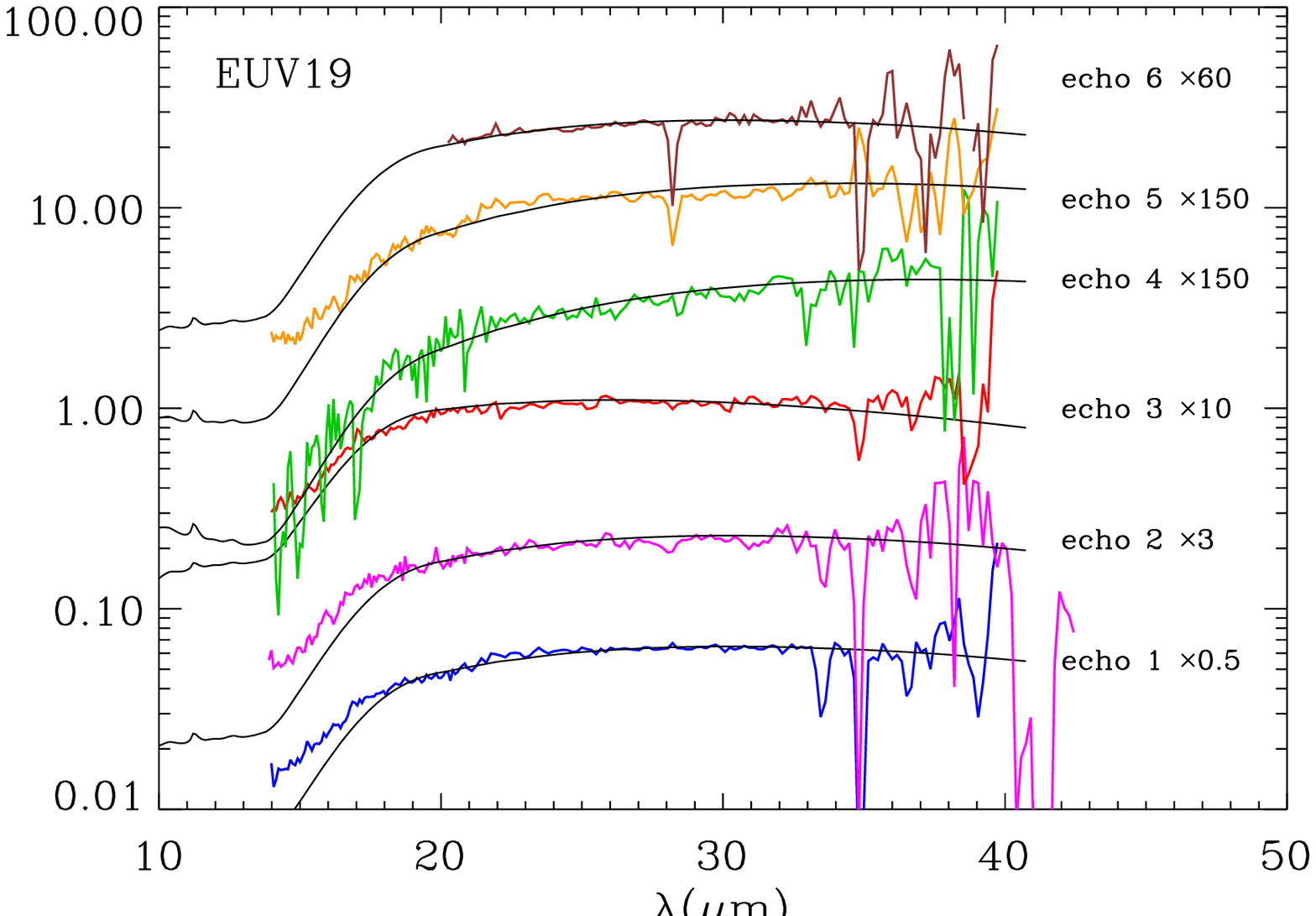}
          \includegraphics[width=2.1in]{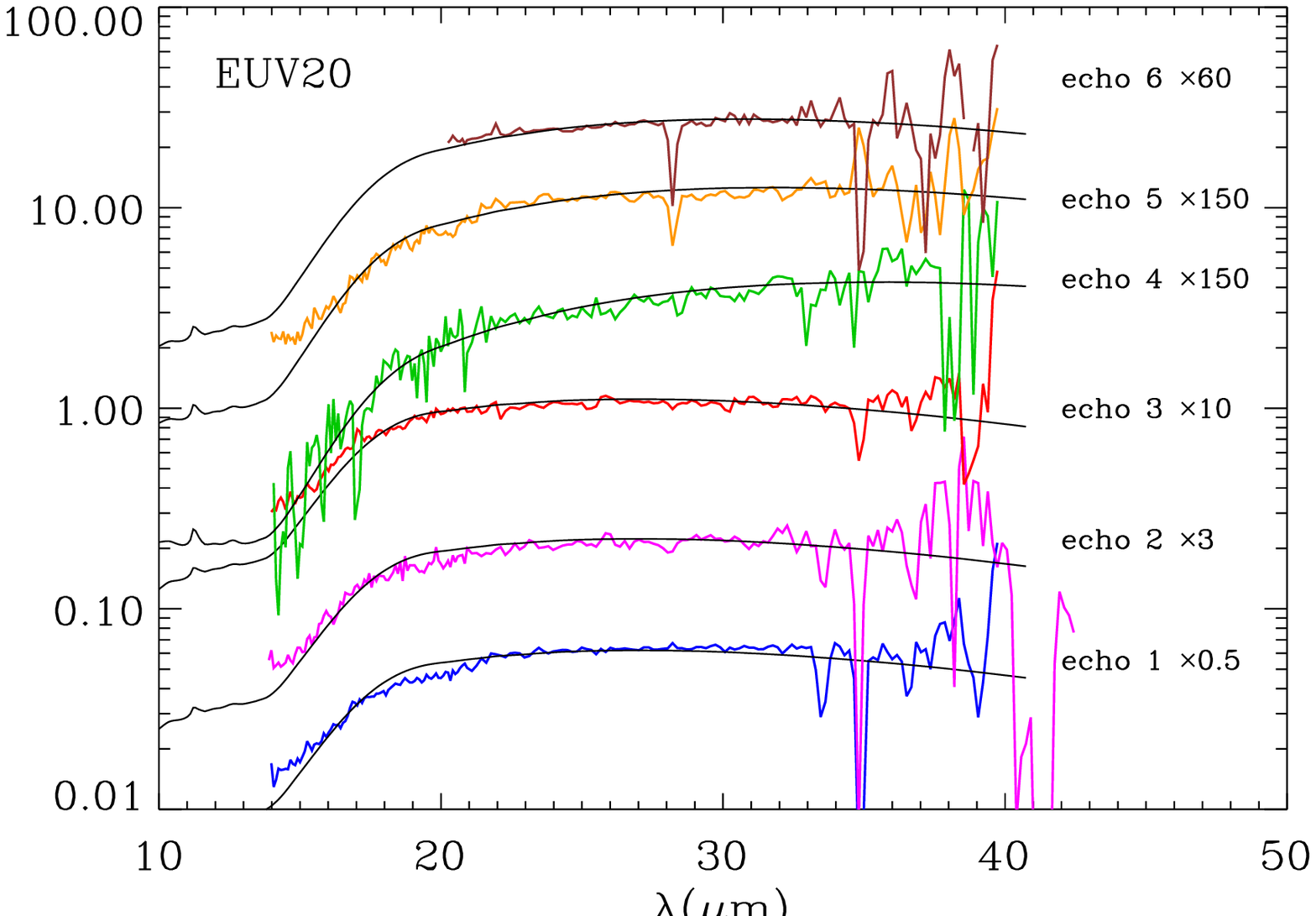} \\
    \vspace{0.1in}
      \includegraphics[width=2.1in]{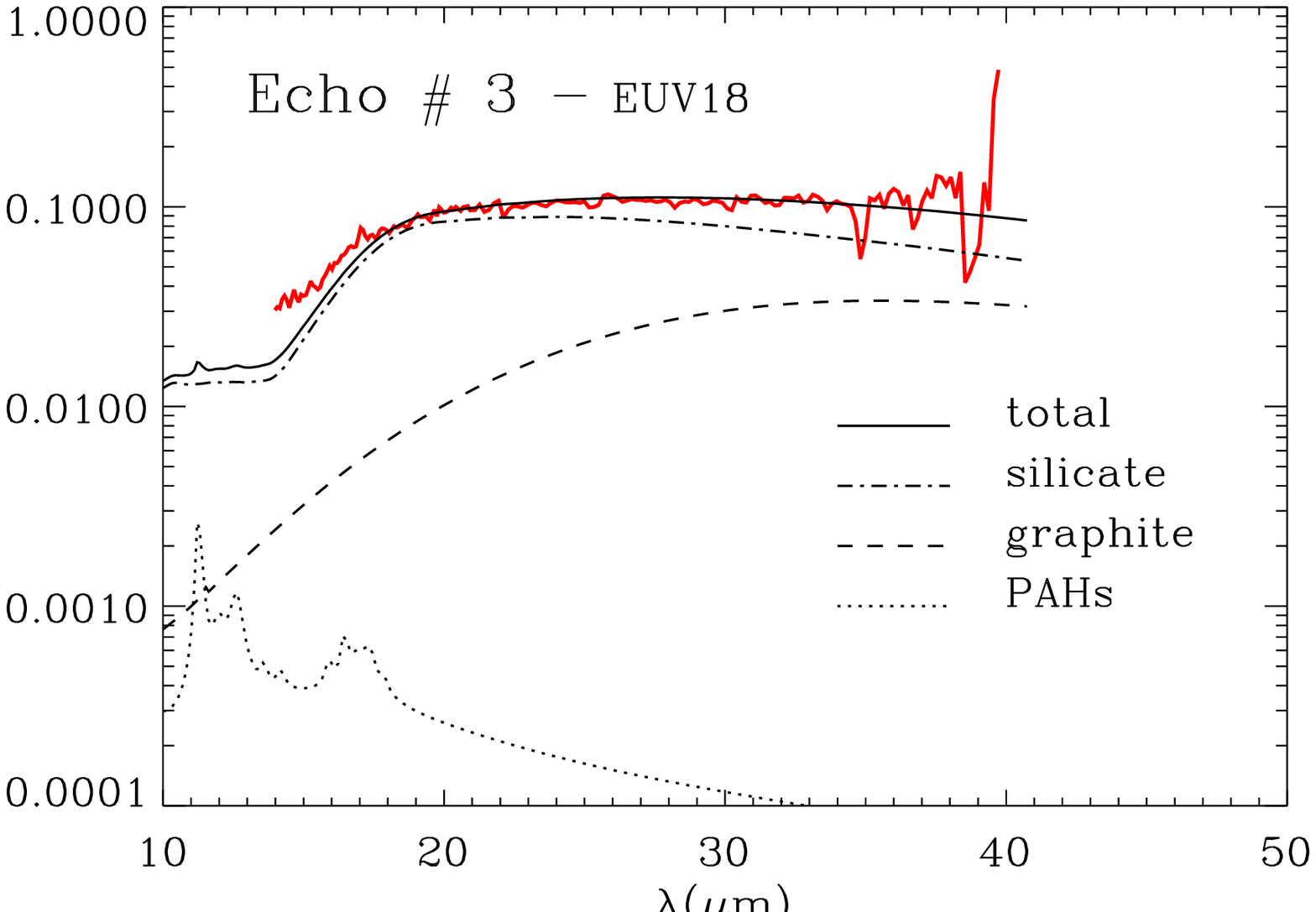}
         \includegraphics[width=2.1in]{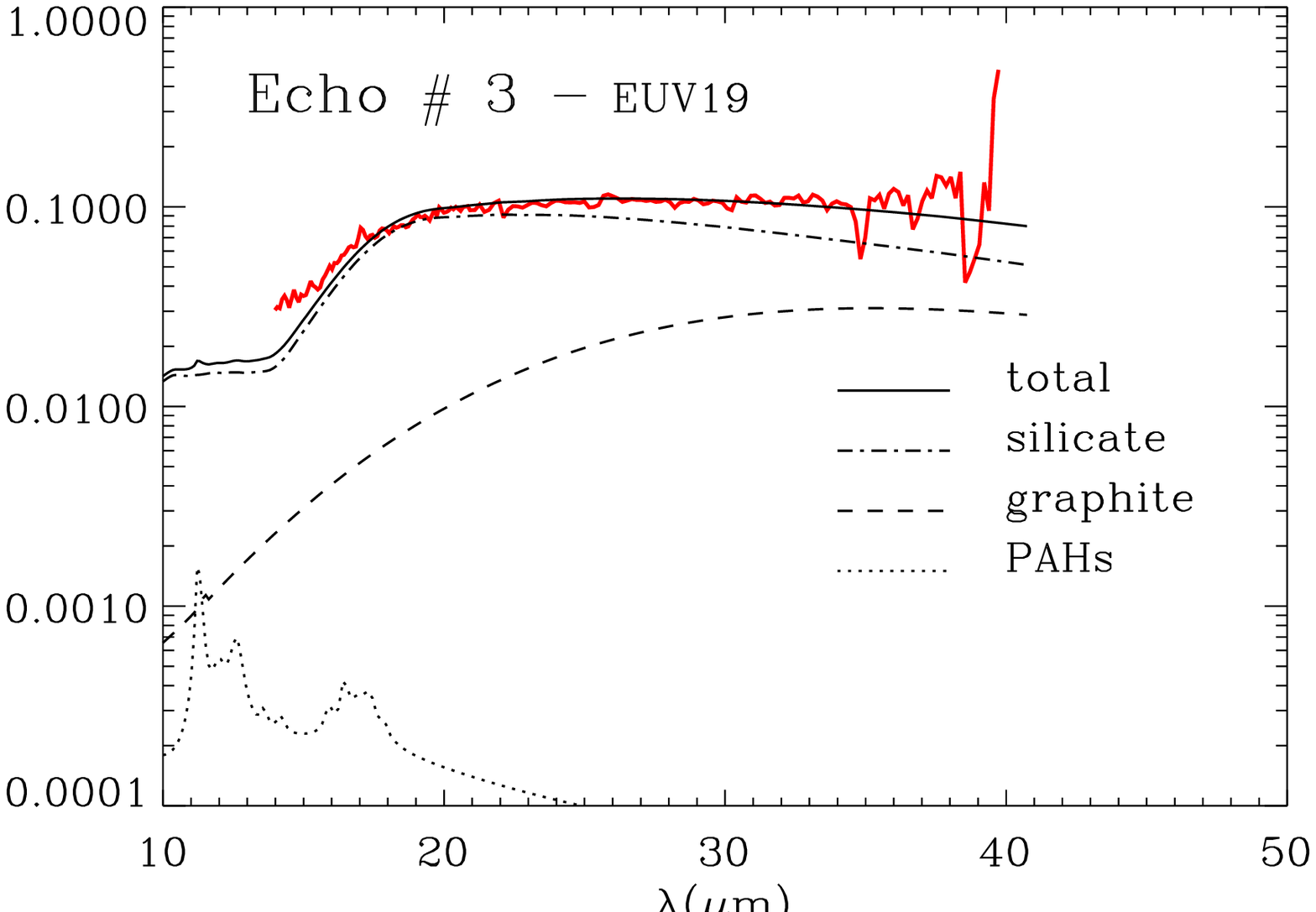}
            \includegraphics[width=2.1in]{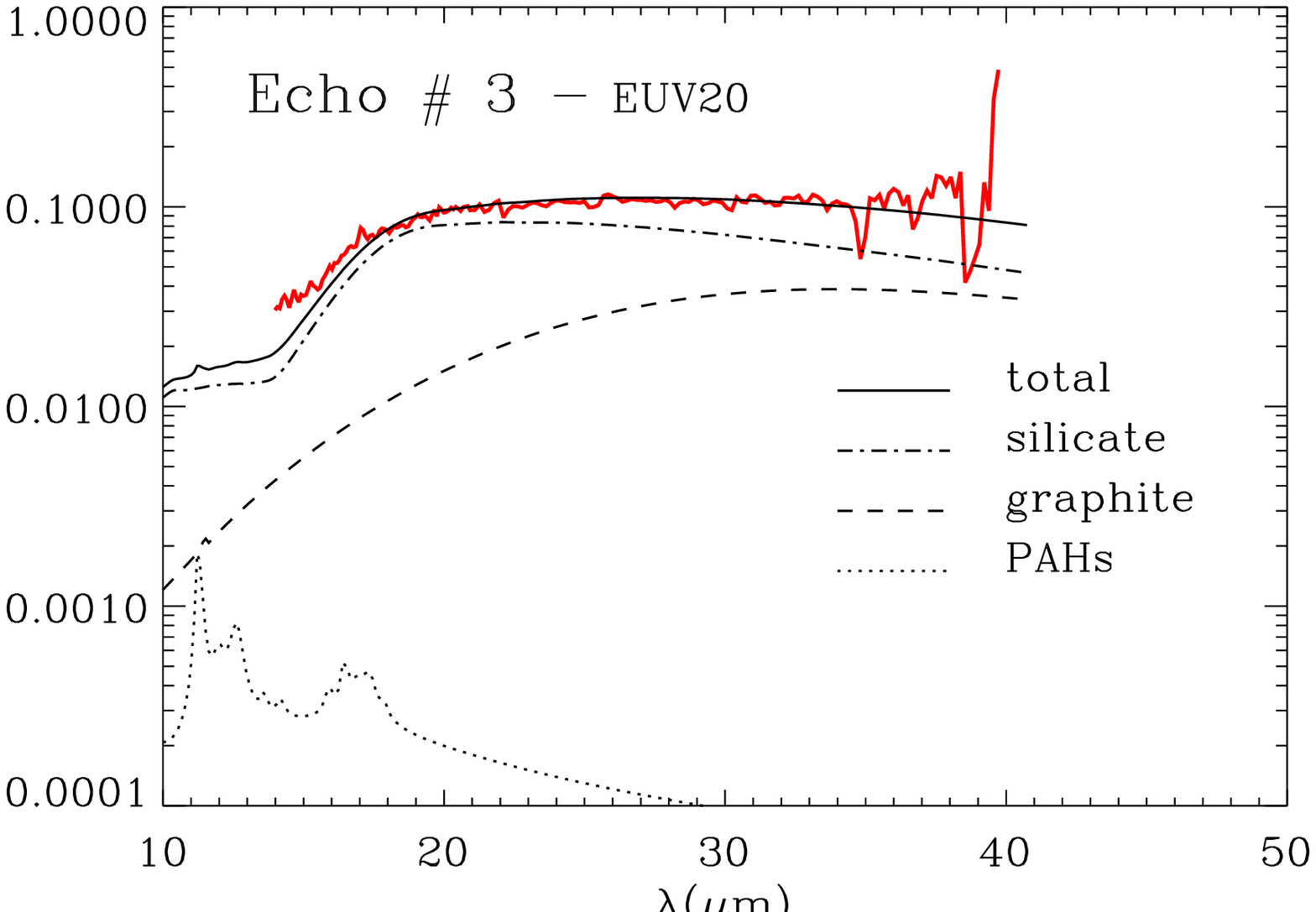}
  \caption{{\footnotesize
Model fits for the EUV burst propagating through a densities of 0.01 (left column), 0.10 (middle column), and 1.0~ cm$^{-3}$ (right column). {\bf Top row:} Calculated IR spectra of an echo exposed to different fluxes characterized by values of $\xi$ = 0.001, 0.008, 0.06, 0.10, 0.40, and 1.0 (left panel); $\xi$ = 0.001, 0.006, 0.02, 0.08, 0.40, and 1.0; and $\xi$= 0.1, 0.4, 0.8, 2.0, 6.0, and 10. All fluxes were calculated for $r_0 = 160$~lyr (see eq. \ref{flux}) and are normalized to unity at $\lambda = 20$~\mic. {\bf Middle row:} Best fit models to the IR echoes. {\bf Bottom row:} Decomposition of the best fitting IR spectrum of Echo~3 into its emission components from silicate and graphite grains and PAHs. A detailed discussion of the figure is in the text.}}
\label{euv}
\end{figure}

\begin{figure}
  \centering
  \vspace{-0.3in}
  \includegraphics[width=2.1in]{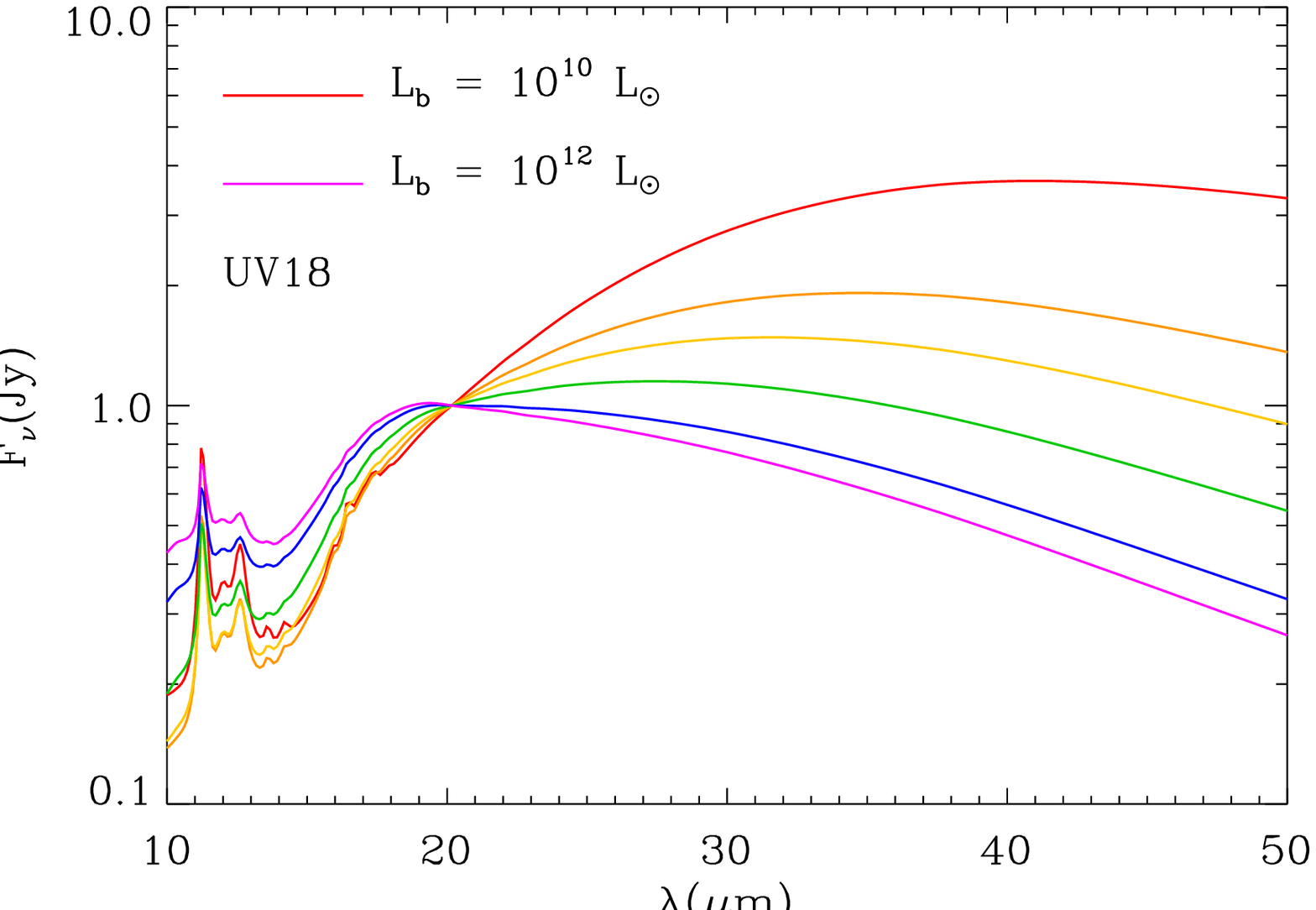}  
   \includegraphics[width=2.1in]{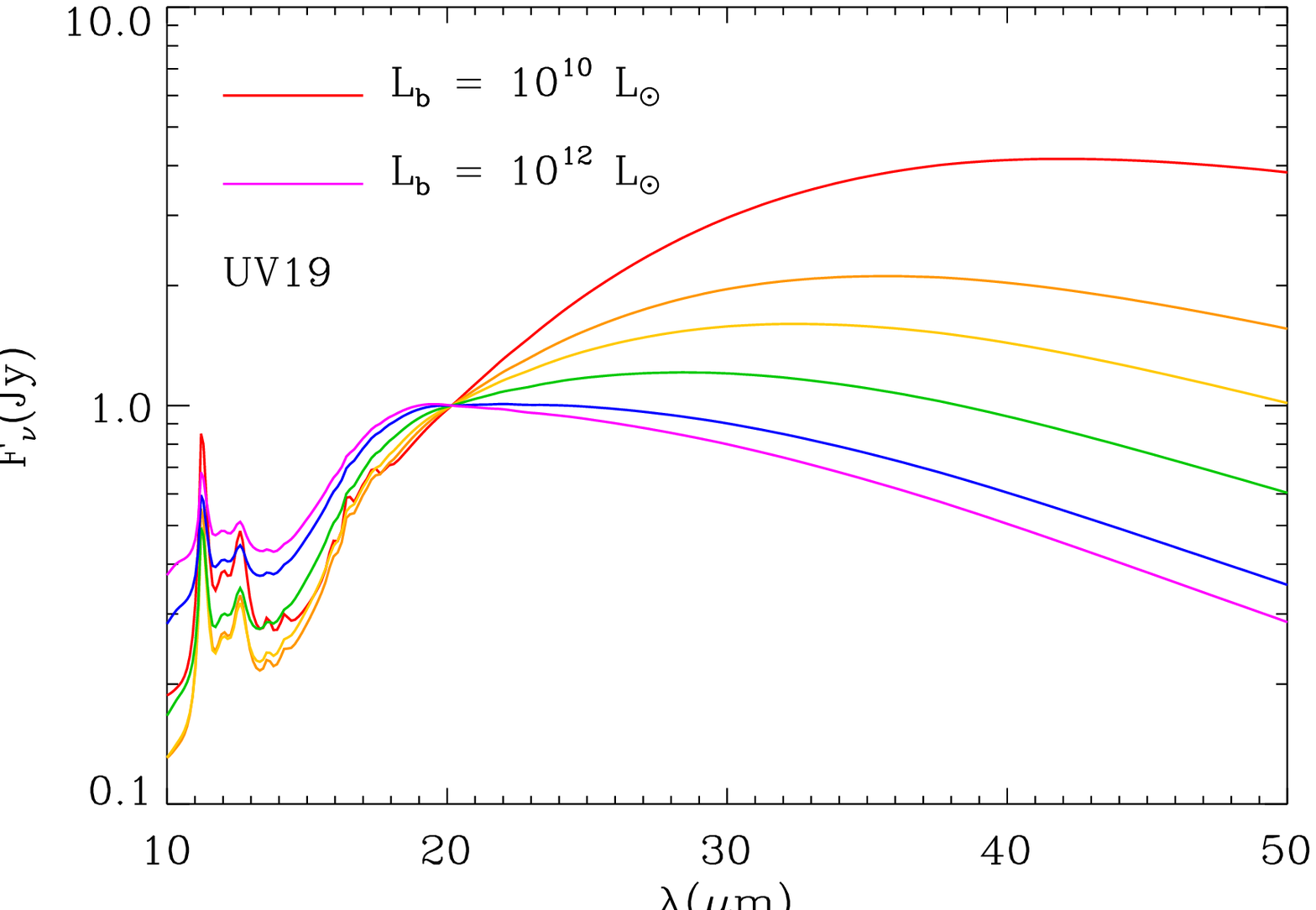} 
    \includegraphics[width=2.1in]{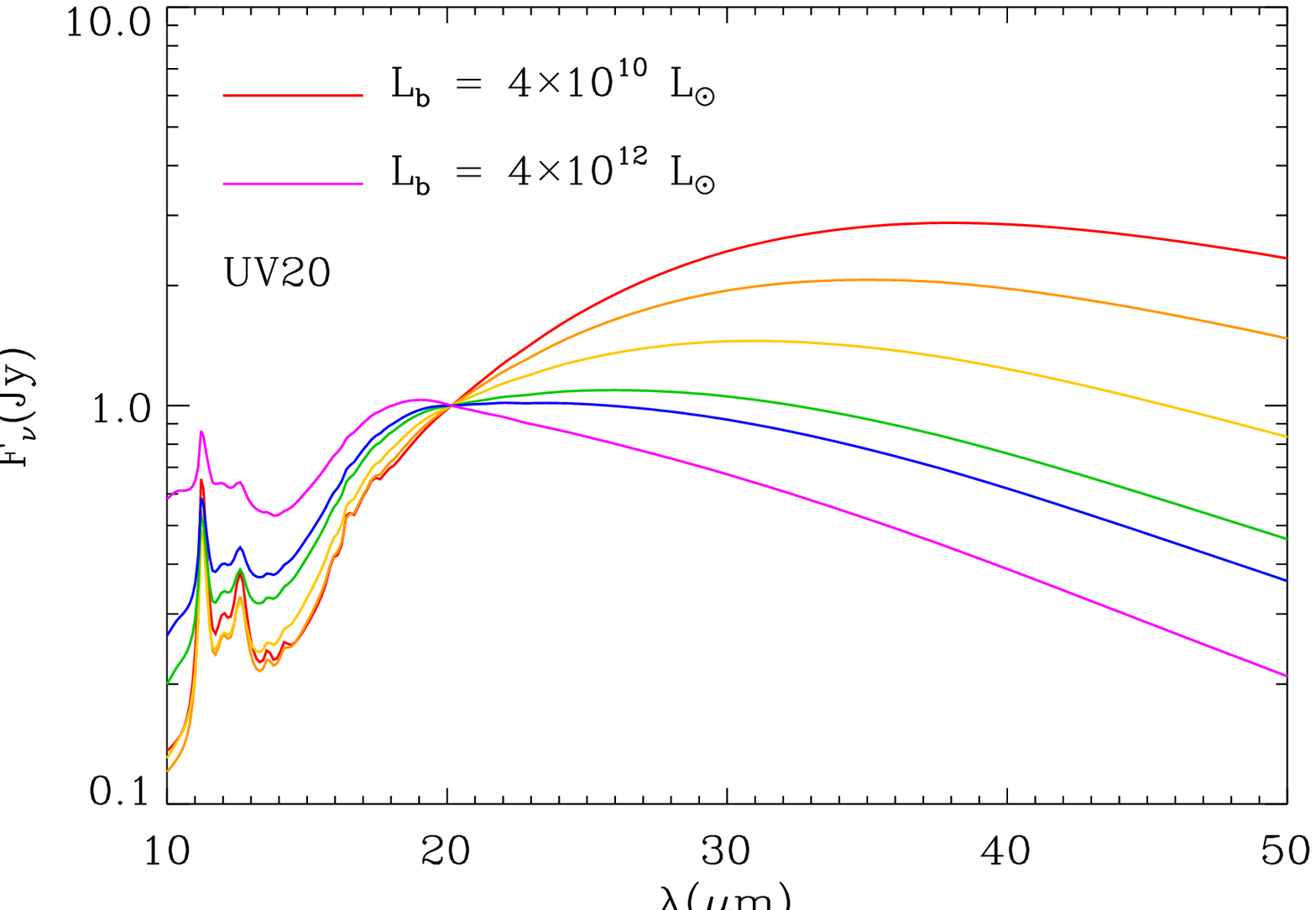} \\
  \vspace{0.1in}
    \includegraphics[width=2.1in]{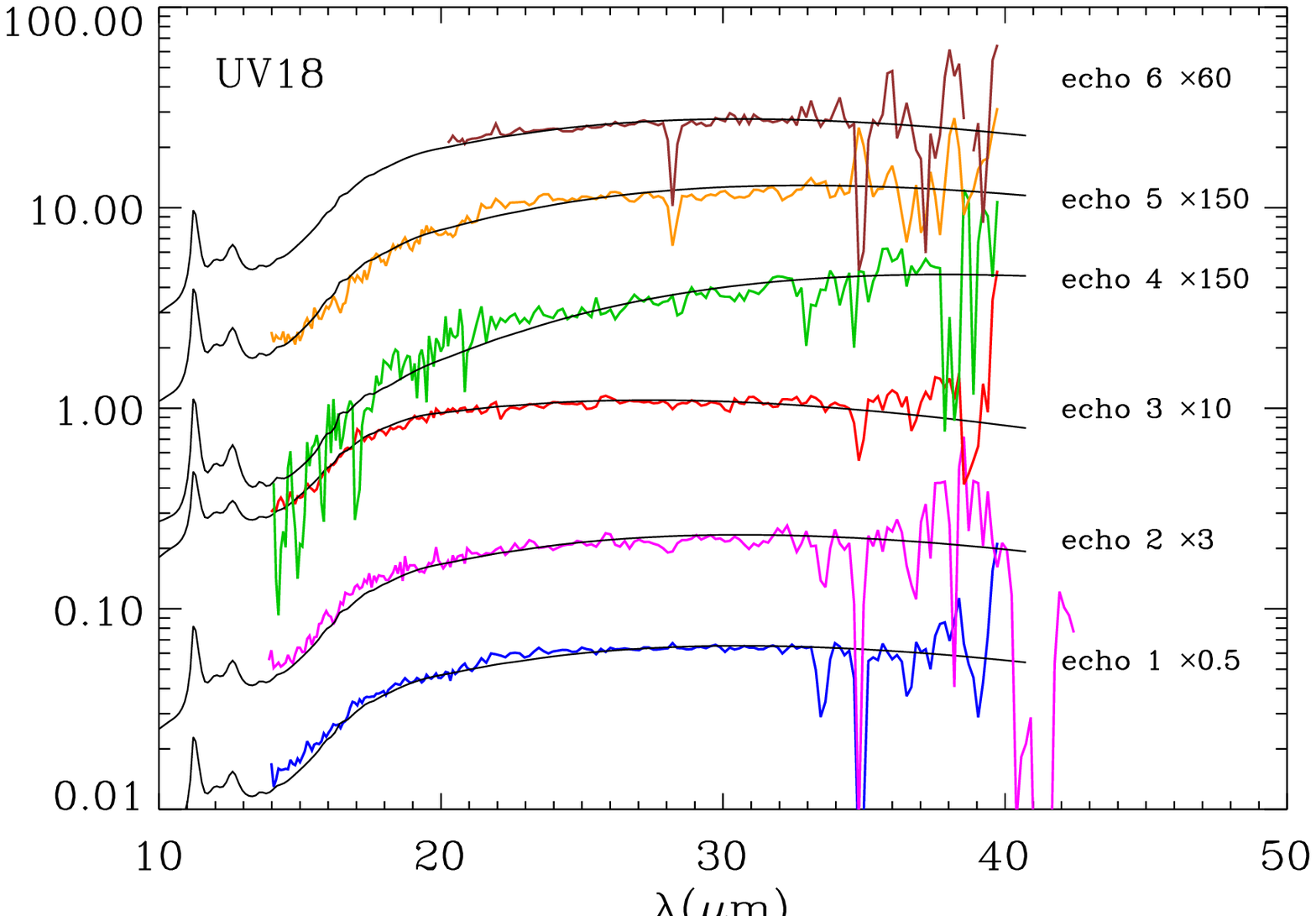}
       \includegraphics[width=2.1in]{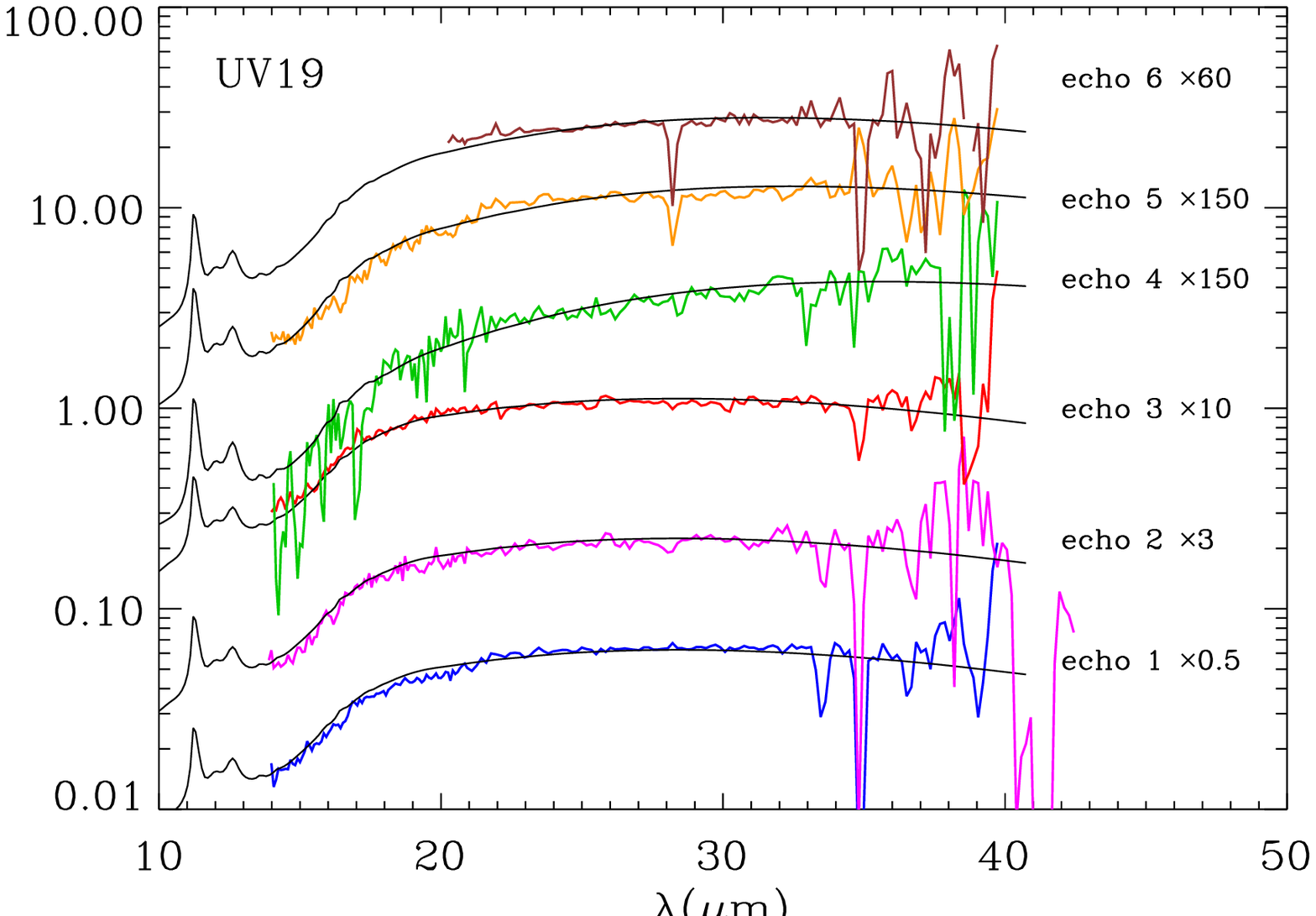}
          \includegraphics[width=2.1in]{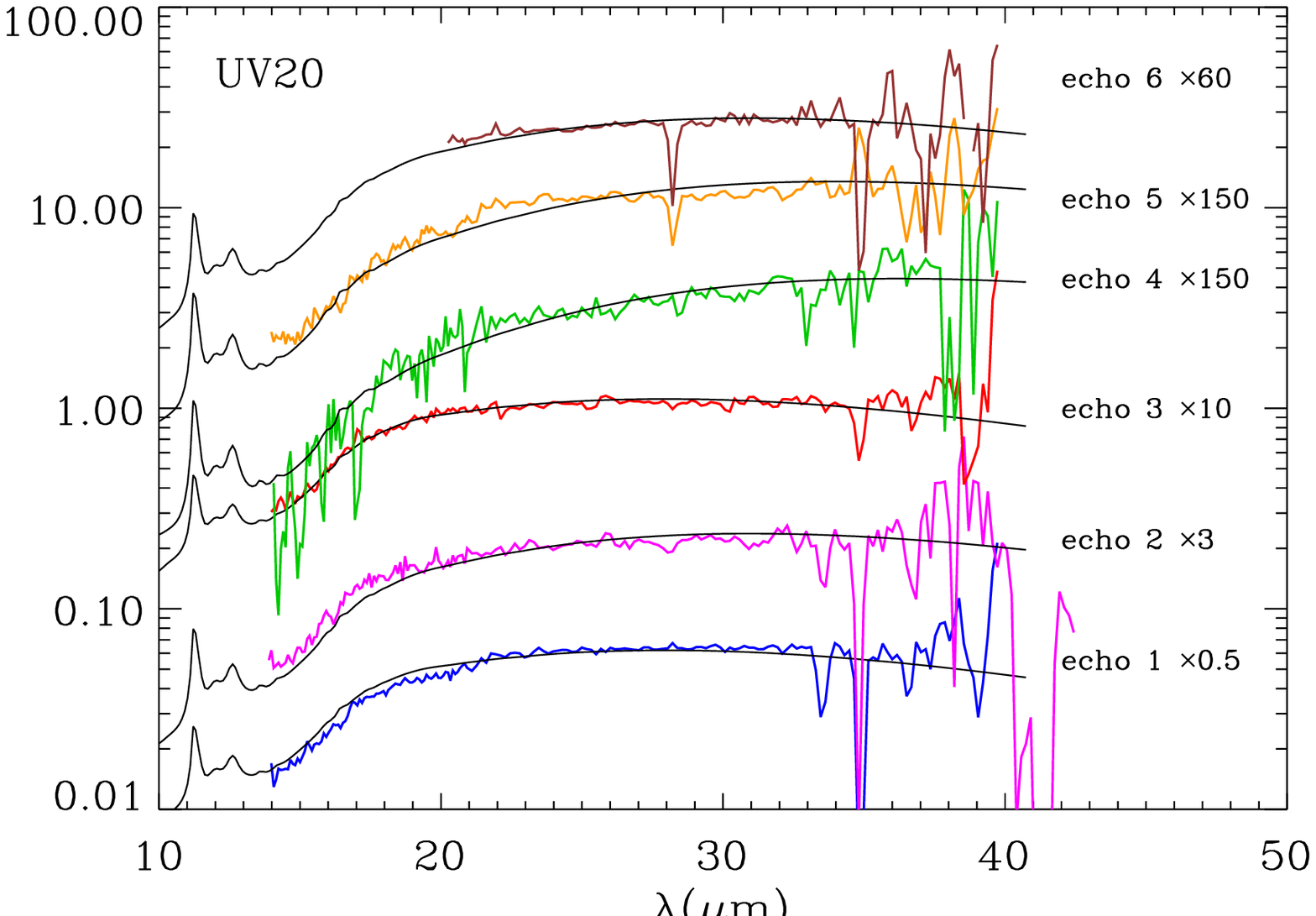} \\
    \vspace{0.1in}
      \includegraphics[width=2.1in]{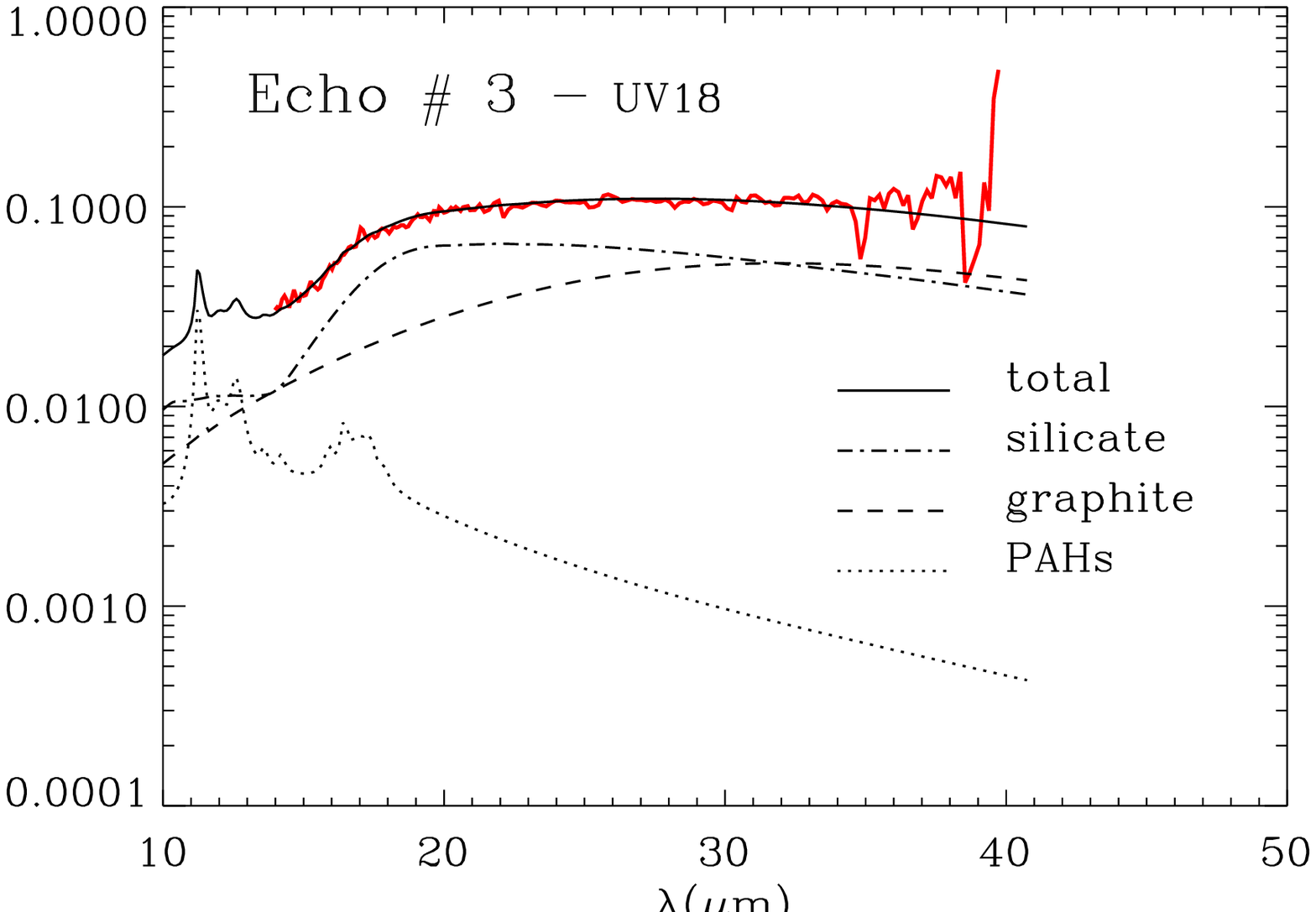}
         \includegraphics[width=2.1in]{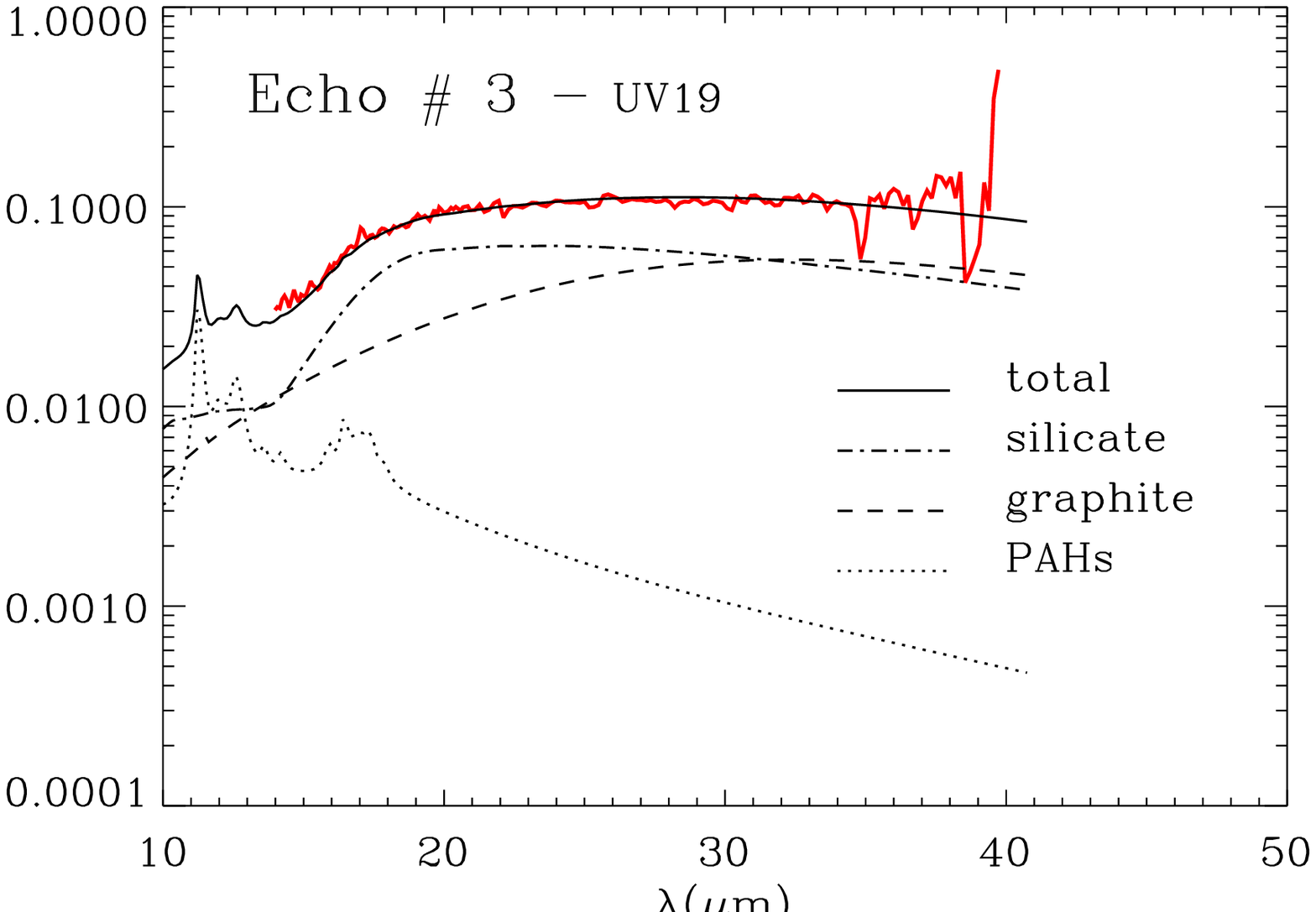}
            \includegraphics[width=2.1in]{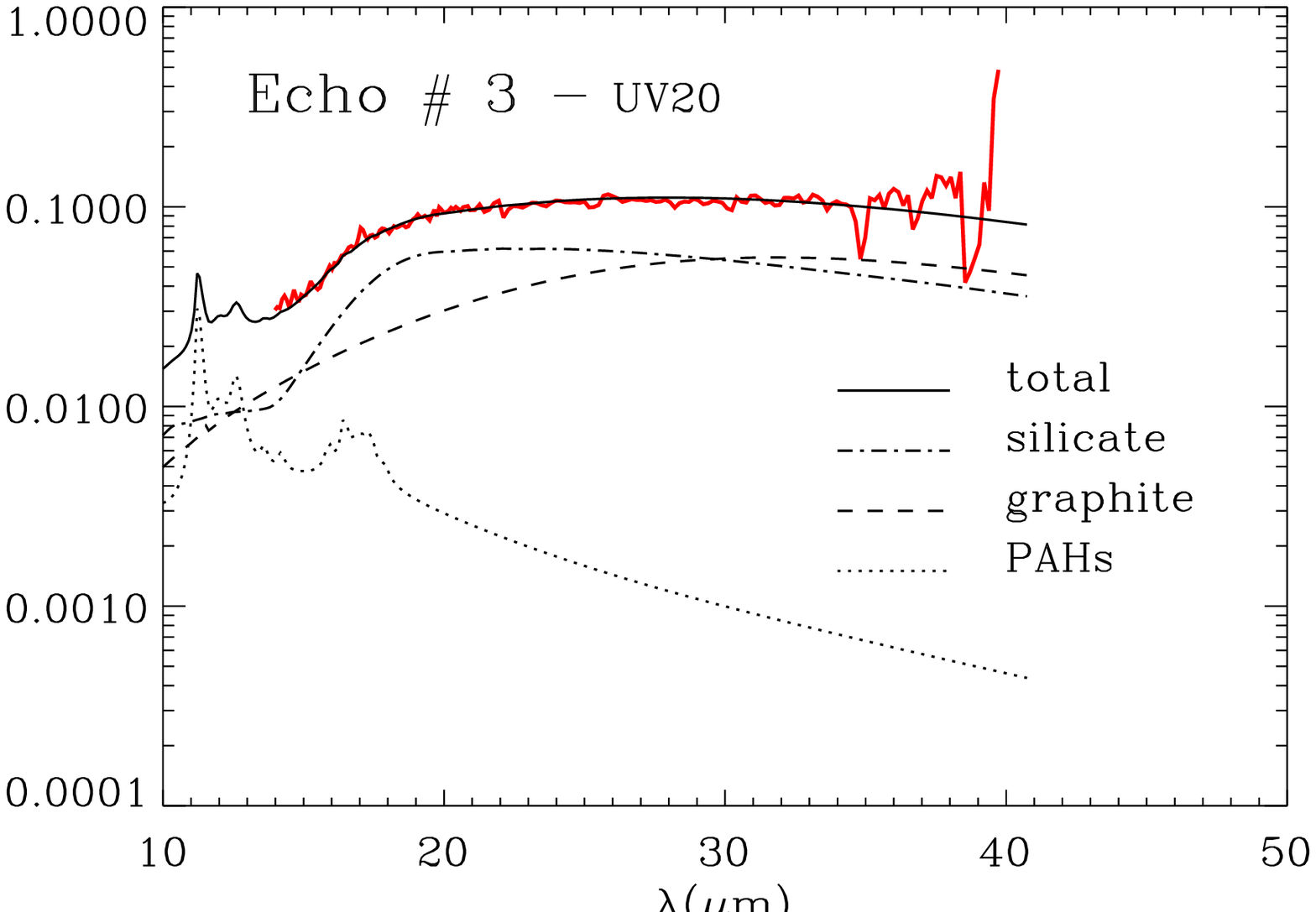}
  \caption{{\footnotesize
The same as Figure \ref{euv} for the UV burst. Calculated IR spectra of an echo exposed to different fluxes characterized by values of $\xi$ = 0.01, 0.04, 0.08, 0.20, 0.60, and 1.0 (left and middle panel); and $\xi$= 0.04, 0.08, 0.20, 0.60, 1.0, and 4.0. All fluxes were calculated for $r_0 = 160$~lyr (see eq. \ref{flux}) and are normalized to unity at $\lambda = 20$~\mic. {\bf Middle row:} Best fit models to the IR echoes. {\bf Bottom row:} Decomposition of the best fitting IR spectrum of Echo~3 into its emission components from silicate and graphite grains and PAHs. A detailed discussion of the figure is in the text.}}
\label{uv}
\end{figure}

Figures \ref{olc}, \ref{euv} and \ref{uv} depict the results for the \olc, the EUV and the UV bursts, respectively. The top panels in the figures show the calculated IR spectra of dust that is exposed to a range of illuminating fluxes, characterized by different  values of $\xi$. To illustrate the dependence of the spectral signature of the dust on the flux and spectral shape of the incident radiation we normalized all fluxes to unity at 20~\mic. 
The middle panels depict the best fits to the echo spectra by the incident fluxes. Different echoes required exposure to different fluxes to fit their IR spectra. The bottom panels of the figures show the contributions of the different dust components to the total flux from Echo~3.

The figures show that the radiation from the optical burst cannot reproduce the observed characteristic rise in the echo spectra from 14 to 20~\mic. In contrast, both the EUV and UV models reproduce the echo spectra fairly well, independent of the density of the intervening medium. 
 
\begin{figure}[htbp]
  \centering
  \vspace{-0.3in}
  \includegraphics[height=2.0in]{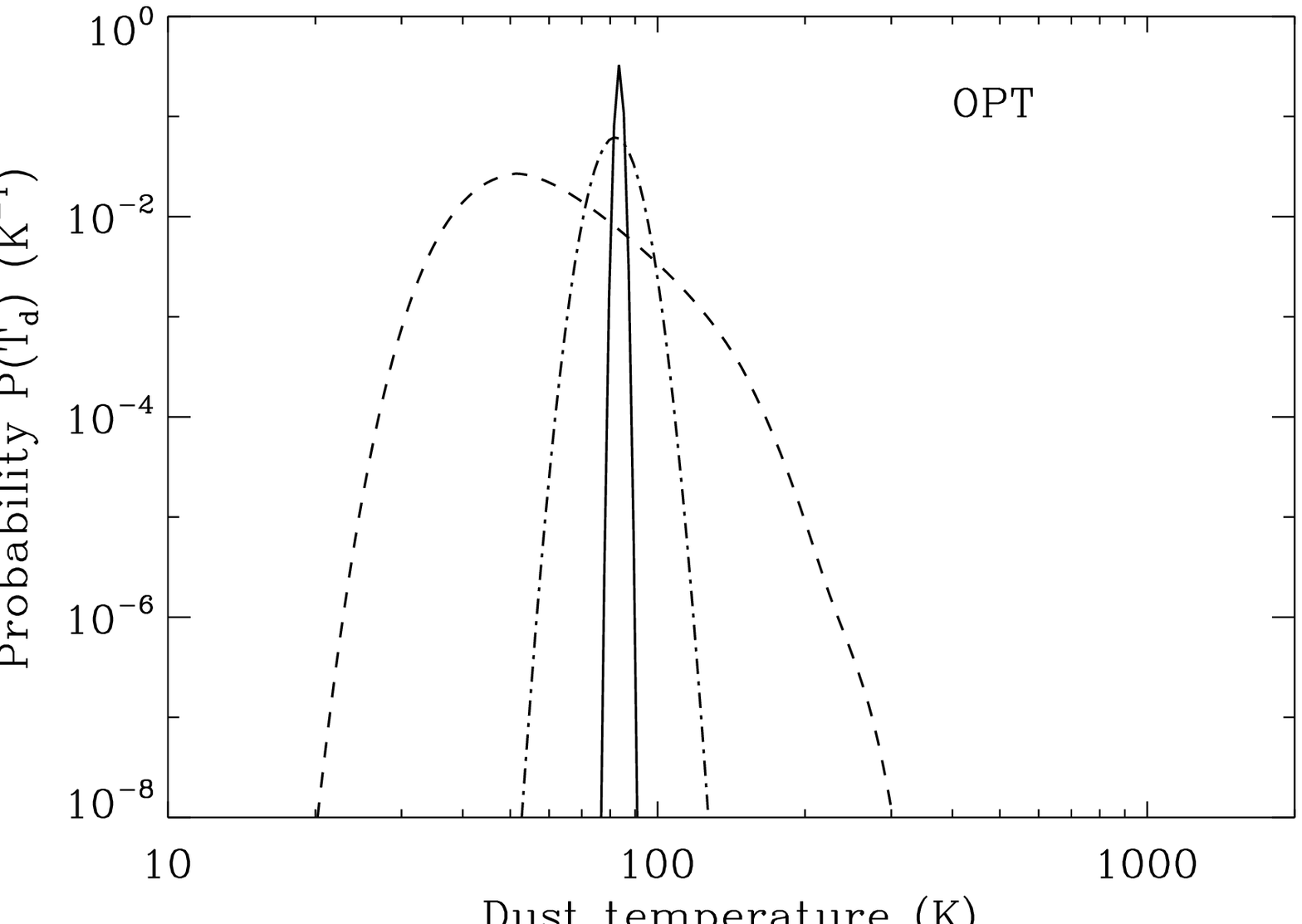} \\
  \vspace{0.1in}
    \includegraphics[height=2.0in]{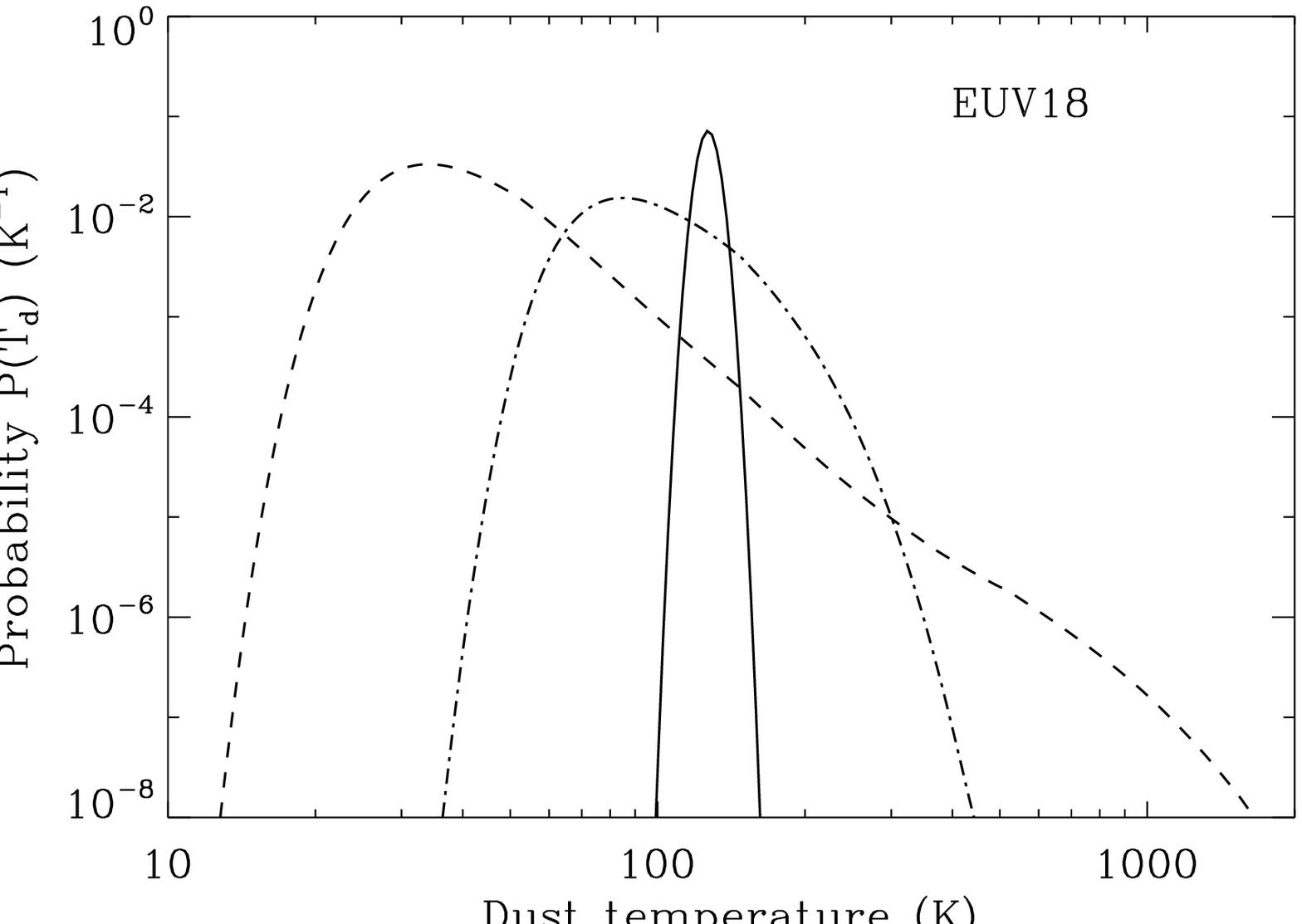} \\
    \vspace{0.1in}
      \includegraphics[height=2.0in]{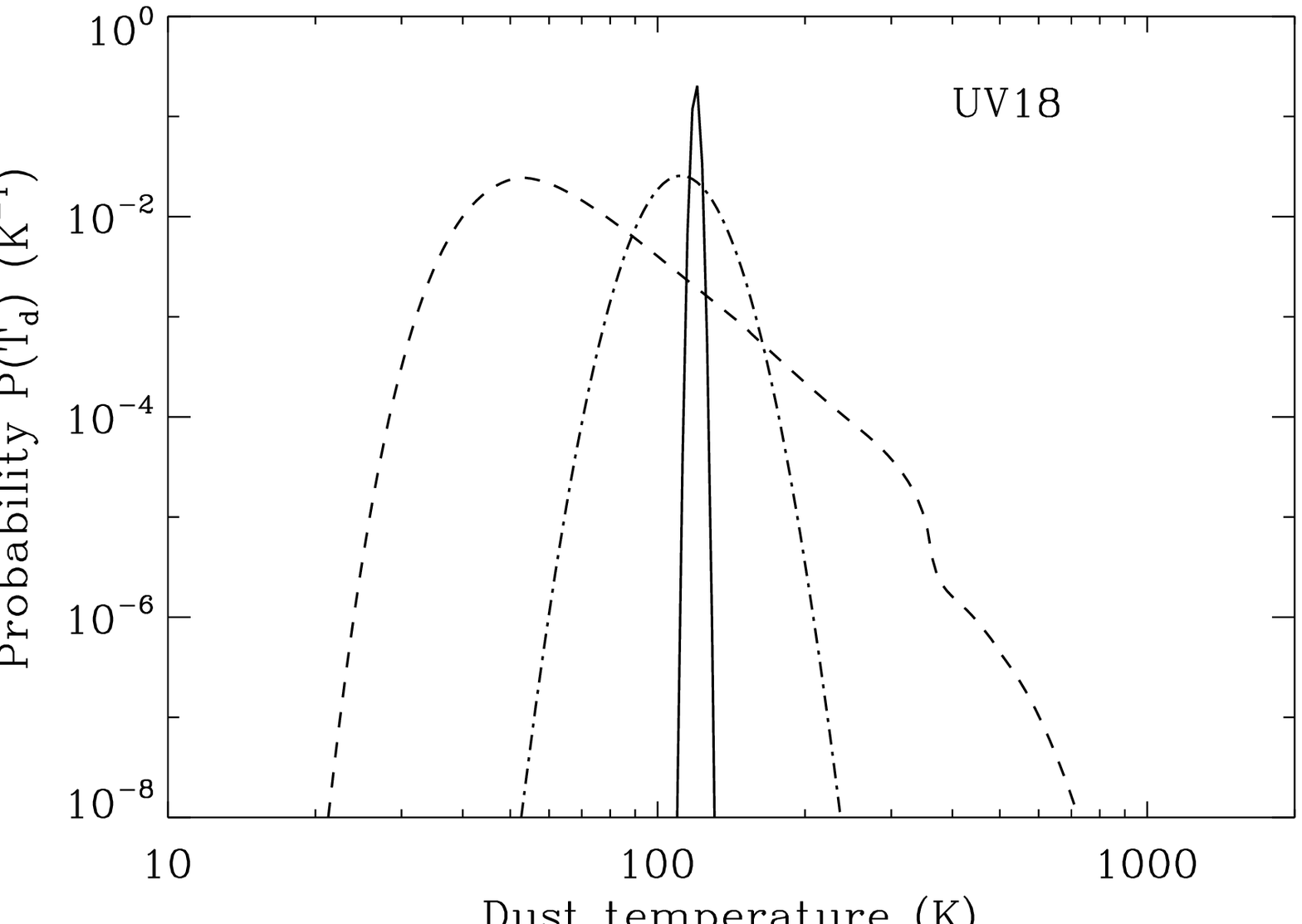}
  \caption{{\footnotesize
The temperature probability distribution $P(T_d)$~(K$^{-1}$) for 0.0010~\mic\ (dashed line); 0.0030~\mic\ (dashed-dotted line), and 0.010~\mic\ (solid line) silicate grains exposed to the optical burst (top panel), the EUV18 burst model (middle panel), and the UV18 burst model (bottom panel) with fluxes calculated for $\xi$ = 0.6, 0.2, and 0.1, respectively [see eq. \ref{flux})]. Larger grains radiate at the equilibrium dust temperature. Only the EUV and UV burst are capable of heating a significant fraction of the silicate grains to the temperatures ($\gtrsim 100$~K) that are needed to produce the 18~\mic\ feature. A more detailed discussion of the figure is in the text.}}
\label{tempfluc}
\end{figure}

The decomposition of the fits to Echo~3 into the various dust components illustrates that the rapid rise in the echo spectra between $\sim 14 - 20$~\mic, is caused by the 18~\mic\ silicate feature. Two conditions have to be met to produce the observed rise in the echoes' spectra: (1) the silicate grains have to be heated to temperatures in excess of $\sim 150$~K; and (2) the silicate emission has to dominate the emission from graphite grains which have a smooth spectrum over the $\sim 10 - 40$~\mic\ wavelength interval. The echoes generated by the optical burst fails to meet both conditions. Silicate grains have a low opacity to optical photons and therefore remain  at relatively low temperatures. The top panel of Figure~\ref{tempfluc} shows that the equilibrium silicate temperature is only about 83~K for the \olc\ burst. Stochastic heating is limited to the smallest grains with sizes $\lesssim 70$~\AA. In  contrast, graphite grains have a much larger opacity at visible wavelengths and attain much higher equilibrium temperatures of $\sim 175$~K. Consequently, the IR echo spectrum produced by the optical burst is dominated by graphite emission.

The EUV and UV bursts meet both requirements for producing the rise in the echo spectra. As shown in Figure~\ref{tempfluc}, they both possess sufficiently hard photons to raise the equilibrium silicate temperature to $\sim 150$~K. Compared to the optical burst, a larger fraction of the small grains are fluctuating to temperatures beyond $\sim 100$~K. Furthermore, the opacity of silicate grains rises dramatically in the EUV, and becomes comparable to that of the graphite grains. Consequently, as illustrated in the bottom row of Figs. \ref{euv} and \ref{uv}, the total IR spectrum is dominated by silicate emission. 

\subsection{A Fit to the Average Echo Spectrum}

\begin{figure}[htbp]
  \centering
  \includegraphics[width=5.0in]{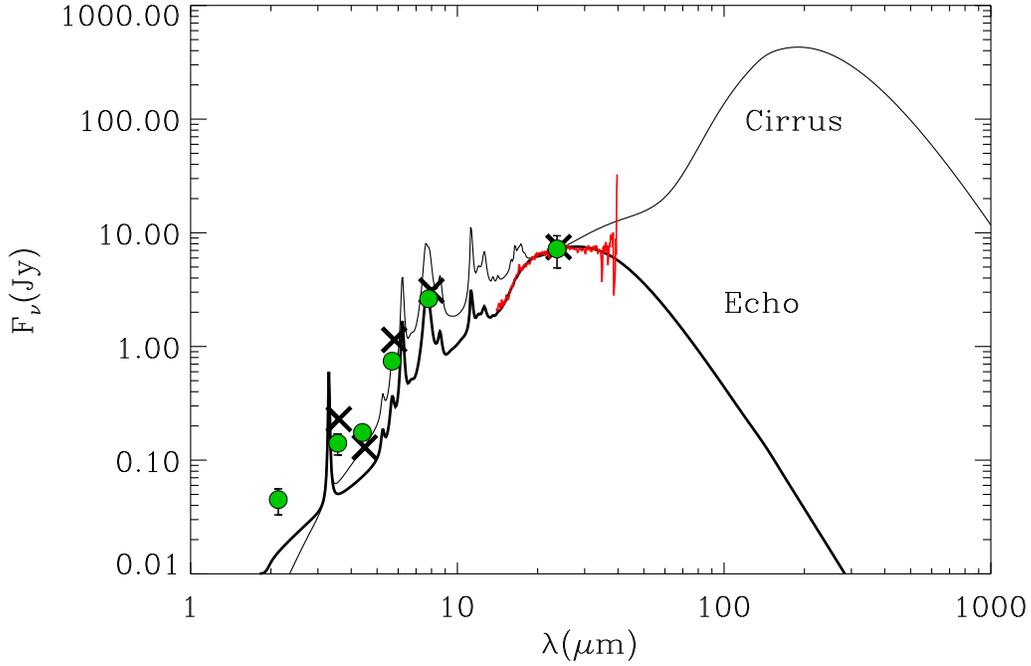} 
   \caption{{\footnotesize
The spectral energy distribution of echoes in the K-band (2.2~\mic), IRAC (3.6, 4.5, 5.8, and 8.5~\mic) and MIPS (24~\mic) bands, averaged over 59 positions in the northern lobe region of the MIPS image  \cite{krause05} is represented by filled green circles. The X symbols in the figure represent our model fit (bold solid line) to the spectrum of echo 3 (red line), convolved with the  IRAC and MIPS 24~\mic\ bands, and normalized to the 24~\mic\ MIPS datum point. The fit to the average echo spectrum required the abundance of PAHs to be increased by a factor of 2 over their general ISM abundance. Also shown in the figure is the average cirrus spectrum \citep{zubko04}. Observed only  in the IRAC and MIPS 24~\mic\ bands, the average echo spectrum is very similar to that of interstellar cirrus. The excess 2.2~\mic\ emission is caused by direct reflection a different component of the SN lightcurve.}}
\label{krause}
\end{figure}

It is of interest to examine to what extent our model fits to the $\sim$14--40~\mic\ echo spectra also fits the average spectral energy distribution (SED) obtained by \cite{krause05} in the 2.2, 3.6, 4.5, 5.8, and 24~\mic\ bands.  
This SED, shown in Figure~\ref{krause}, was obtained by averaging over 59 position in the northern lobe region of the preliminary MIPS image around the remnant. The bold line in the figure shows the model fit to the IRS spectrum of echo 3 (orange line). The X symbols represent this model's flux densities after integration over the IRAC and MIPS 24~\mic\ bandpasses. These flux densities were normalized to fit the average echo flux density at 24~\mic. The fit to the average echo spectrum required the abundance of PAHs to be increased by a factor of 2 over their general ISM abundance. The increased PAH abundance did not affect the fit to the spectrum of echo 3 since PAHs contribute relatively little emission at wavelengths $\gtrsim 14$~\mic. Also shown in the figure is the average cirrus spectrum \citep{zubko04}.
 
The figure shows that if the echoes are only observed in the broad IRAC and MIPS 24~\mic\ bands, their colors are  very similar to those interstellar cirrus dust. Consequently, the true temperature of the echoes is totally lost with this limited photometric data. In contrast, the \iras\ 25-to-60~\mic\ color ratio was clearly able to distinguish between the echoing clouds and the interstellar cirrus (see Fig.~\ref{casa_iris}), confirming the unique spectral signature of the echoes. The temperature of the echoing dust is much higher than that of the dust in cirrus clouds, requiring  an incident flux that is significantly harder than the general  ISRF. 

The excess 2.2~\mic\ emission is caused by reflection, and may therefore sample a totally different portion of the SN light curve, powered by the radioactivitis in the ejecta. The intensity of the reflected echo could therefore provide valuable information of the  Cas~A progenitor. The column density of  the scattering dust could therefore be significantly higher than that  of the thermally emitting dust, providing a geometry-dependent observational test of this hypothesis.
  
\section{MODEL RESULTS}

\subsection{Burst Scenario and Derived Burst Properties}
The incident flux that provides the best fit to a given echo spectra can be converted to burst luminosity by:
\begin{equation}
\label{lburst}
L_b = \left({r\over r_0}\right)^2\, \xi_F\, L_0
\end{equation} 
where $r$ is the {\it actual} distance of the particular echo from the SN, $L_0 = 1\times 10^{12}$~\lsun, and \xsi\ is the value of $\xi$ that produced the best fitting incident flux. 

Each echo requires a particular incident flux to produce its IR spectrum. Given this flux, the value of \xsi\ depends on the assumed column density of the intervening ISM [see eq. (\ref{flux})]. A higher column density will require larger values of \xsi, corresponding to larger burst luminosities.

\begin{figure}
  \centering
  \vspace{-0.3in}
   \includegraphics[width=3.in]{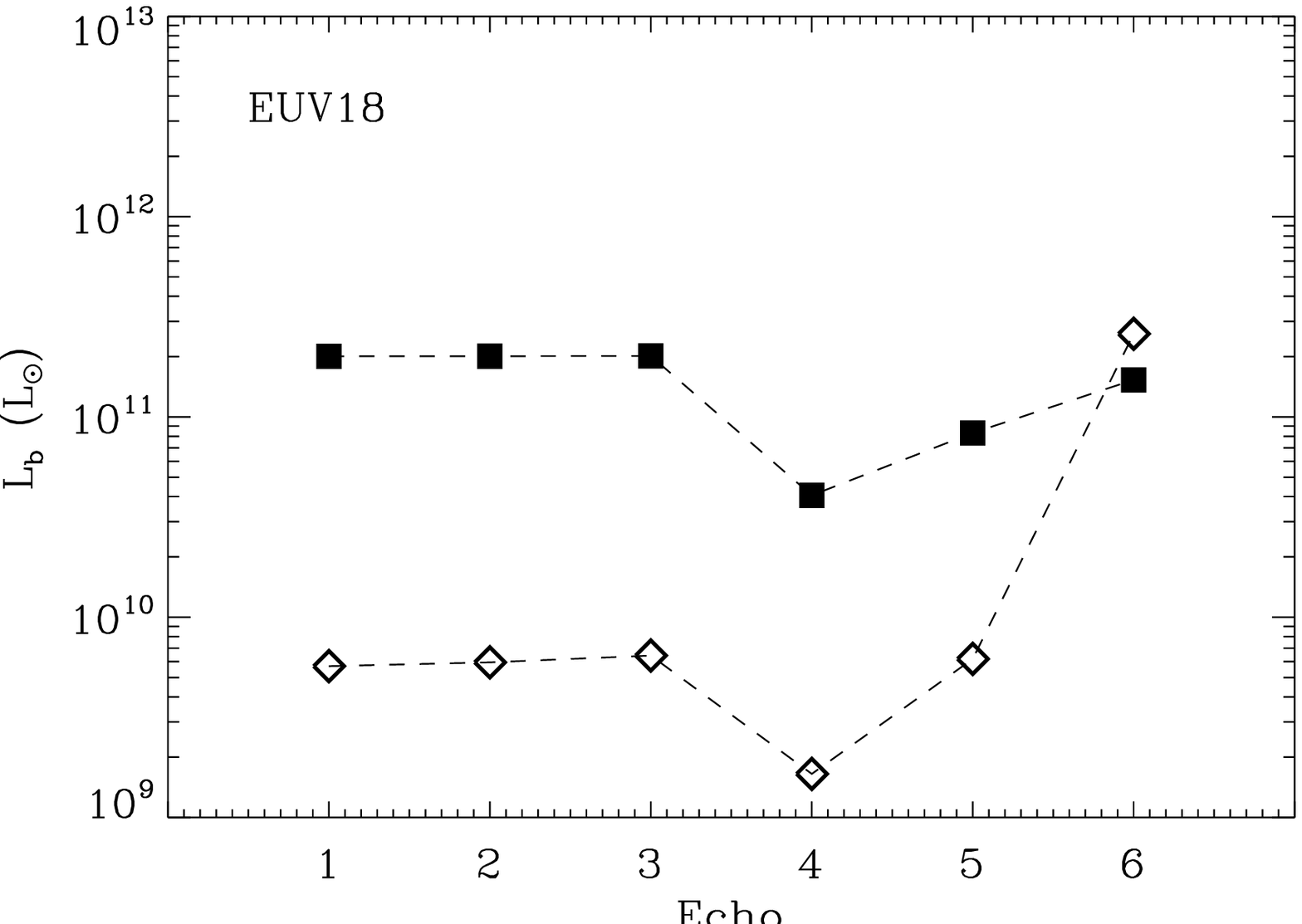} \\
  \includegraphics[width=3.in]{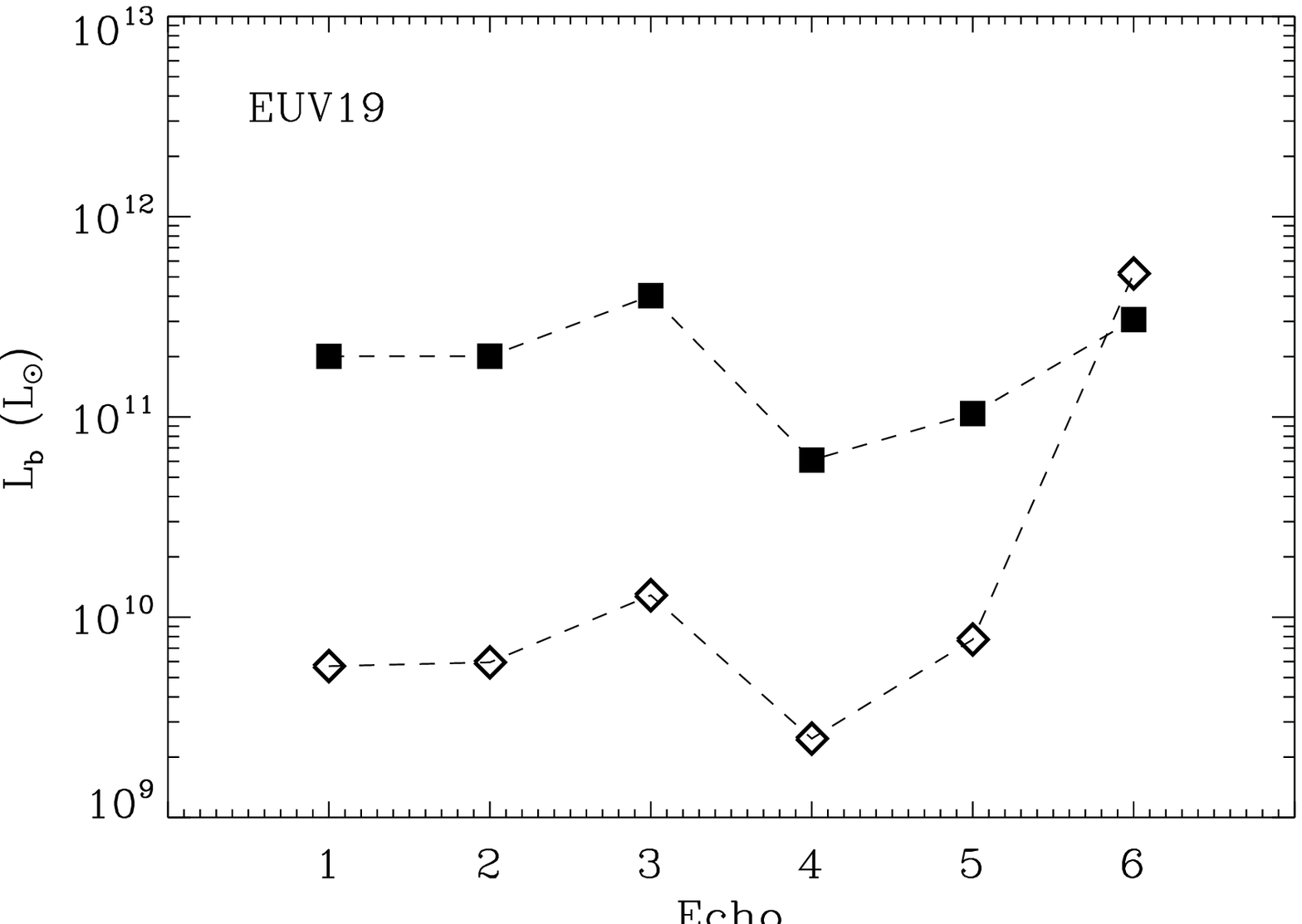} \\
   \includegraphics[width=3.in]{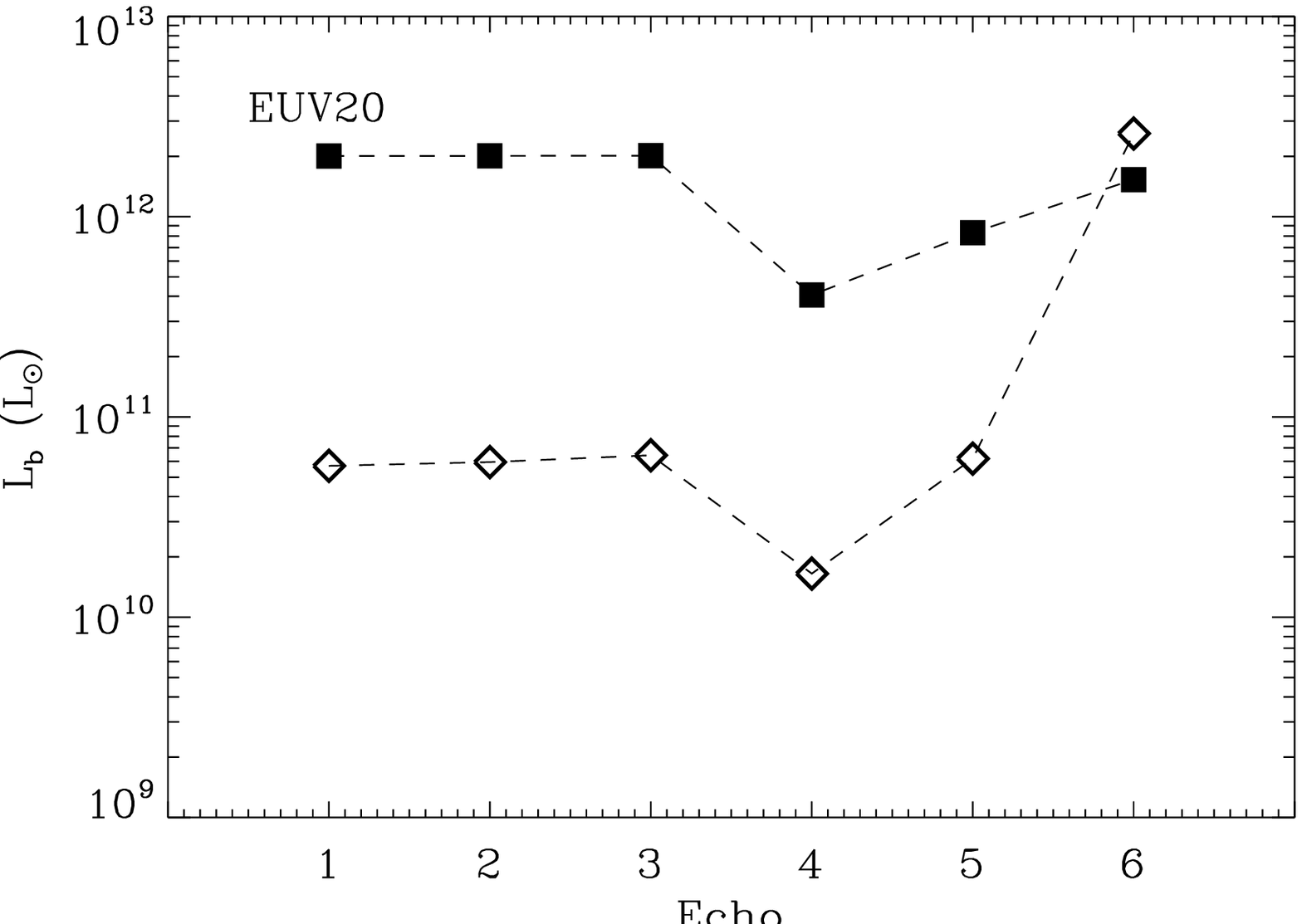}  
    \caption{{\footnotesize
The EUV burst luminosities required to generate the IR echoes from Cas~A [see eq. (\ref{lburst})] for different values of the density of the ISM around Cas~A: $n_H = 0.01$~cm$^{-3}$ (top panel); $n_H = 0.1$~cm$^{-3}$ (middle panel); and $n_H = 1.0$~cm$^{-3}$ (bottom panel). Filled squares are values calculated for a delay time of 320~yr, whereas the open diamonds are calculated for a delay time of 50~yr.
A viable model should give a single value for \lburst\ for all echoing clouds. The scenario in which the delay time is 360~yr is more physical, since it produces a significantly narrower range of burst luminosities compared to the 50~yr model.
The spread in burst luminosities in t=320~yr scenario may reflect the variations in the burst attenuation en route to the echoing cloud. 
}}
\label{lumburstEUV}
\end{figure}

\begin{figure}
  \centering
  \vspace{-0.3in}
   \includegraphics[width=3.in]{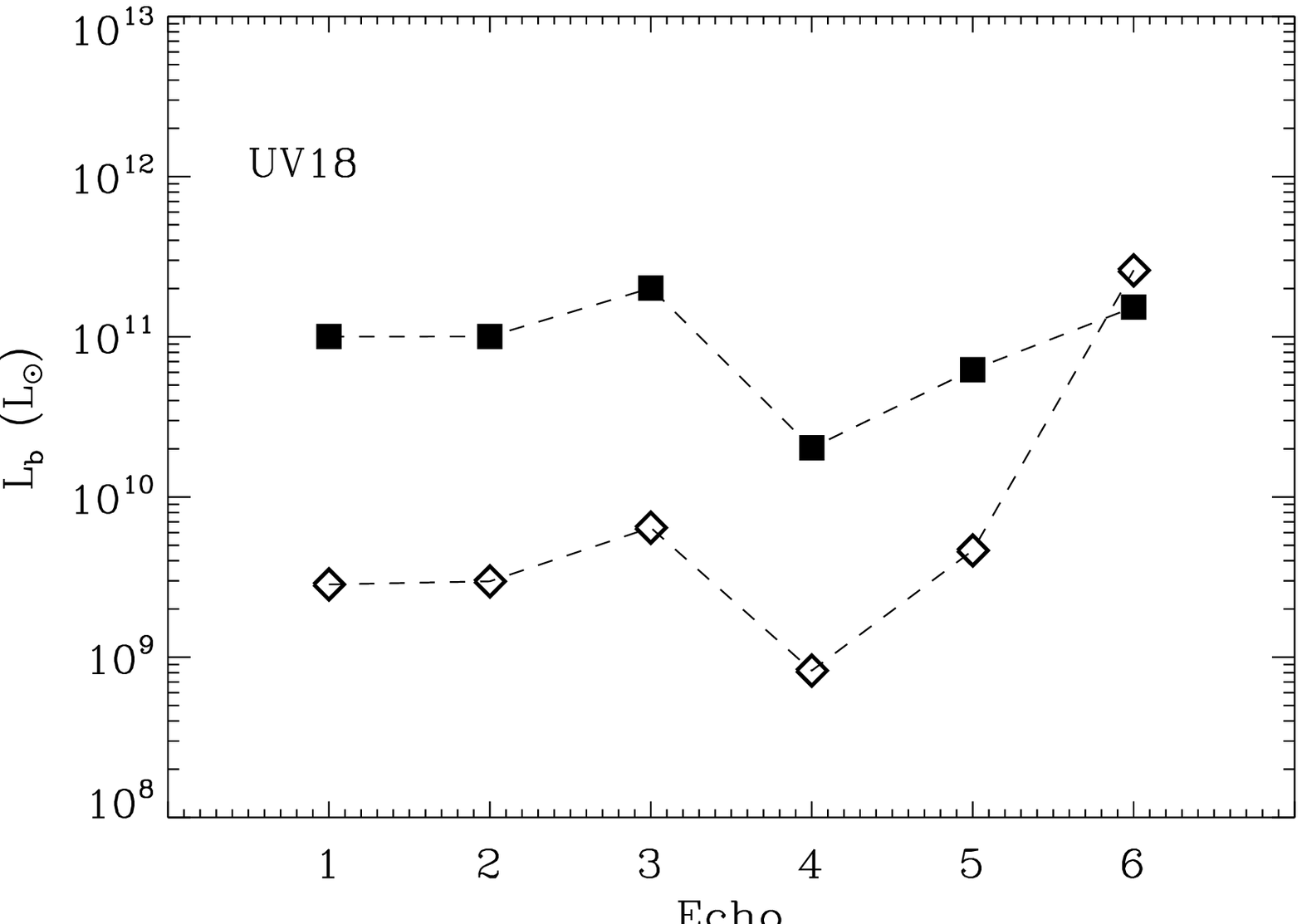} \\
  \includegraphics[width=3.in]{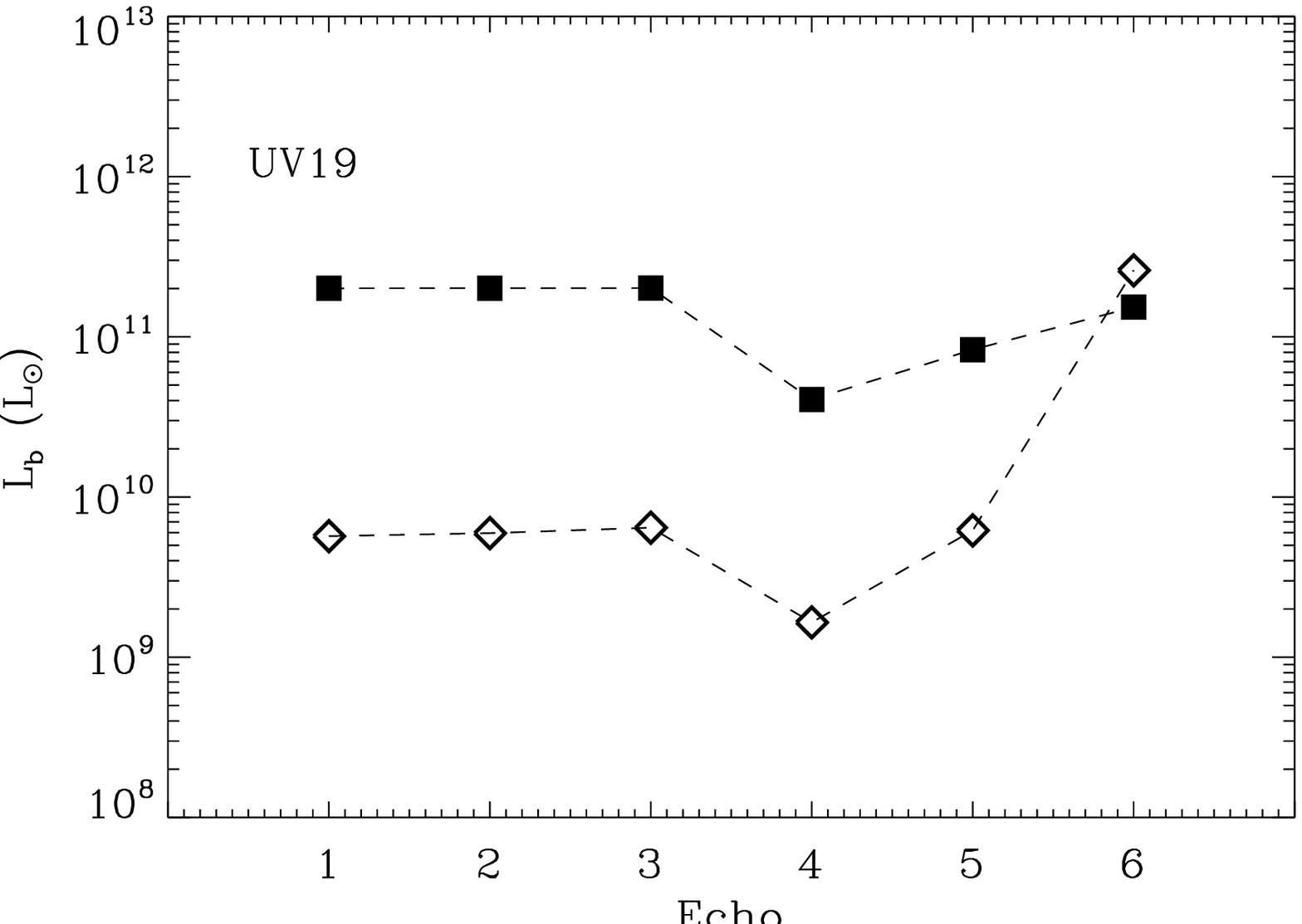} \\
   \includegraphics[width=3.in]{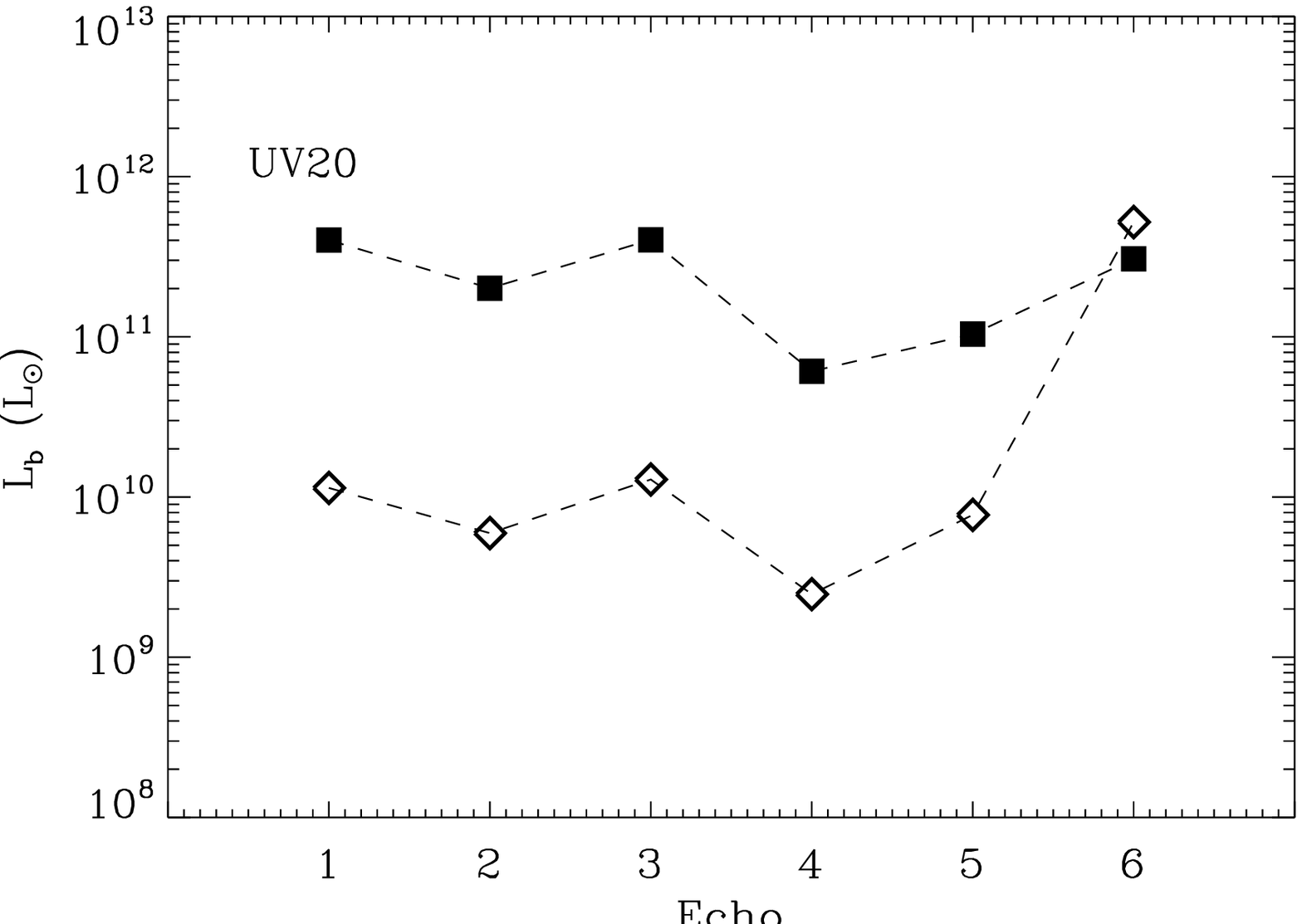}  
  
  \caption{{\footnotesize
Same as Figure \ref{lumburstEUV} for the UV burst.
}}
\label{lumburstUV}
\end{figure}

Figures~\ref{lumburstEUV} and \ref{lumburstUV} depict the burst luminosities required to produce the echo spectra for the different burst models and delay time scenarios. We assume that all echoes were produced by a single burst. A viable echo scenario should therefore give the same burst luminosity for all echoes. The results show that a scenario with the 50~yr delay time produces a large spread in burst luminosities, compared to the 320~yr delay scenario. The figures shows that the dispersion in the derived burst luminosities for the latter scenario is only about 50\% of the mean value (see also Table 3), compared to a factor of two in the former. This is a direct consequence of the fact that the echo spectra are generally very similar, thus requiring exposure to the same incident flux of radiation. Narrowing the range of burst luminosities for the 50~yr delay model will require more distant echoes to have systematically lower column density, a non-physical and contrived solution to this problem. We therefore support the conclusion of \cite{kim08} that the echoes were not generated by a hypothetical burst from the neutron star occurring about 50~yr ago. In  principle, all echoes could have been produced by repeated bursts occurring  over a period of 320~yrs. However, a more likely scenario is that the all echoes were produced by a single burst associated with the Cas~A SN event.    

Table 2 lists the model fits for each of the echoes assuming a 320~yr delay time. For each model we present the incident flux required to produce the best fit to the echo spectrum as characterized by the value of \xsi, the burst luminosity, $L_b$, the column density of the echoing cloud, $N_H$, and the $\chi^2$ of the fit. For each echo, the value of $\chi^2$ was normalized to unity for the best fitting model. 
From the values of the \chisq\ of the fit, we see that the \olc\ burst provides a significantly worse fit to the echo spectra than the EUV and UV bursts. 
The EUV and UV bursts are generally of similar quality, although for echoes 1, 2, and 3, UV models are distinctly  better. The models with a lower H-column density usually provided slightly better fits than the models with $N_H = 1.5\times 10^{20}$~cm$^{-2}$. 

The lack of any evidence for H-recombination lines from the gas around Cas~A \citep{krause05} suggests that the burst ran out of ionizing photons before it reached the echoing clouds. Inspection of Figure~\ref{burst_spec} shows that  these conditions are easily met when the intervening H-column density is sufficiently high with $N_H \gtrsim 10^{19}$~cm$^{-2}$. 
   
Table 3 lists the burst luminosity averaged over all echoes for each burst model. It clearly shows the correlation between the derived burst luminosity  and the column density of the intervening ISM. Larger column densities require higher burst luminosities to produce the same flux needed to produce the observed echo spectrum. Excluding model results for the highest column density, we get  that the average EUV-UV burst luminosity is $L_b = 1.5\times10^{11}$\lsun.

\subsection{The EUV-UV Burst Must be Powered by Shock Energy}
In principle, the EUV-UV luminosity could be either powered by the explosion energy or by the radioactive decay energy in the ejecta. A simple analysis, based on the best  estimate of the properties of the Cas~A progenitor \citep{young06}, firmly rules out the latter possibility. 
Based on explosion calculations, ejecta velocity and mass constraints, and $^{44}$Ti yields, \cite{young06} estimate the mass of $^{56}$Ni to be between 0.06 and 0.2 \msun. Taking the average decay energy to be 1.73~MeV and a decay lifetime of 8.8~d, the maximum  luminosity that can be powered by the decay of $^{44}$Ti is between 1.2 and $4.1\times10^9$~\lsun. This luminosity is about two orders of magnitude smaller than that required to produce the IR echoes.

The luminosities required to produce the echoes are easily obtained by tapping into the kinetic energy of the explosion. Taking an average explosion energy of $2\times10^{51}$~erg \citep{young06} suggests that this energy must be released during a period of 40~d to produce an average luminosity of $1.5\times10^{11}$~\lsun.
Several hydrodynamic models for the explosion of SN1987~A were constructed to fit the earliest UV and optical light curve \citep{woosley88,ensman92,blinnikov00}. The models of \cite{blinnikov00} show that the burst had a peak blackbody temperature of $\sim 10^6$~K, cooling on a timescale of days to a blackbody temperature of $\sim 6000$~K, the H-recombination temperature in the expanding photosphere. Burst durations are therefore typically $\lesssim$~days.
Taking a burst decay time of one day, suggests that only 2.5\% of the kinetic energy of the explosion need to be thermalized to produce the required EUV-UV luminosity.  
These simple arguments conclusively show that the echoes have revealed the breakout of the SN shock through the stellar surface. The combined constraints on the energy of the explosion, the required burst luminosity, and hardness of its spectrum can provide useful constraints for  modeling the structure of the progenitor of Cas~A \citep{matzner99}.     

\subsection{Derived Cloud Properties}
From the fits of the calculated IR spectra to the observations we can derive the column density, ${\cal M}_d$ of the emitting dust. We take $\Omega$ to be the width of the IRS LL slit squared, $\Omega=(10.6'')^2$. The corresponding H-column density is given by: $N_H = {\cal M}_d /(m_H\, Z_{dH})$, where $Z_{dH} = 0.0087$ is the dust-to-H mass ratio of the BARE-SG-S \cite{zubko04} interstellar dust model, and $m_H$ is the mass of the H-atom. The column densities derived for each echo and burst model are listed in Table 2. Echo~6 is the brightest, and consequently has the highest H-column density. 
Table 4 lists the column density of each echo, averaged over all EUV and UV burst models. 
The average column density for Echoes 1-5 is about $4.5\times 10^{17}$~cm$^{-2}$, and higher by a factor of $\sim 5$ for Echo~6.

All echoes within $50'$ of Cas~A are located behind the SN. 
The IR emission arises from a narrow region of thickness $c\,$\taub, where \taub\ is the burst duration. along the observer's line of sight that is illuminated by the burst of radiation. The width of the radiation front is determined by the burst duration \taub. The hydrogen number density, $n_H$, of the echoing cloud  can be derived from the dust mass column density, ${\cal M}_d$, by:
\begin{equation}
\label{nh}
n_H  =   \left({{\cal M}_d \over m_H\, Z_{dH}}\right)\, (c\,\Delta \tau_b)^{-1} \approx  {385\over\Delta \tau_b(d)} \ {\rm cm}^{-3}
\end{equation}
where the numerical value was derived for a nominal echo column density of $N_H=1\times 10^{18}\ {\rm cm}^{-2}$.

The burst duration should be taken as the length of time during which the burst spectrum is dominated by EUV-UV emission. From the models of \cite{blinnikov00}, we take this time to be about 1~d, giving an average cloud density of $n_H \approx 385$~\cc.
This is a plausible density for the echoing clouds, corresponding to those of typical cirrus clouds \citep{wolfire03}. Cas~A lies on the far side of the Perseus arm [e.g. \cite{bieging91}], so there are relatively few (if any) denser molecular clouds behind the remnant \citep{ungerechts00} to reflect the radiative burst from the SN.
 
\section{SUMMARY}
The \spitz\ satellite discovered a series of IR echoes around Cas~A that are caused by the delayed arrival of thermal emission from dust heated by the radiative output from the SN. Echo spectra are distinct from the IR emission from the diffuse ISM, exhibiting a sharp rise at wavelengths from $\sim$ 14 to 20~\mic. In hindsight, the echoes can also be found in the \iras\ images from 1983.

We assume that the echoes are generated by thermal emission from interstellar dust heated by a single burst of radiation released after the SN event. We calculated the fluxes incident on the dust required to generate the observed echo spectra for different bursts characterized by blackbody spectra with temperatures of $5\times10^5$ (EUV burst), $5\times10^4$ (UV burst), and 6000~K (optical burst). 
 
Our models show that the rise in the echo spectra is due to the 18~\mic\ silicate feature.  The silicate grains  radiate mostly at the equilibrium temperature of $\sim 120$~K and dominate the echo spectra. The optical burst  is incapable of heating the silicates to sufficiently high temperature required to reproduce the spectral characteristics of the  echoes, and is therefore ruled out as the source of the echoes.

The echoes are instead generated by the intense EUV-UV burst of radiation generated by the shock breakout.
We show that  the echo spectra can be reproduced by bursts with a wide range of temperatures from $\sim 5\times 10^4$ to $\sim5\times 10^5$~K. The burst luminosity required to generate the echoes depends on the assumed H-column density between  the SN  and the echoing cloud. We find  the average burst luminosity to be $\sim 1.5\times 10^{11}$~\lsun\ for an average intervening H-column density of $\sim 1.5\times 10^{19}$~cm$^{-2}$. 

The average H-column density of the echoing clouds is $\sim 5\times 10^{17}$~cm$^{-2}$,  which for a burst duration time of $\sim 1$~d gives an average cloud density of $\sim 200$~\cc. This density is typical of dense cirrus clouds, and consistent with the location of Cas~A behind the Perseus arm and the paucity of dense molecular material behind the remnant

The luminosity required to produce the echoes cannot be generated by the radioactive decay of the 0.06-0.2~\msun\ of $^{56}$Ni that has formed in the explosion, and must therefore reflect the fraction of the kinetic energy of the explosion that was converted to EUV-UV radiation as the SN shock broke out through the stellar surface.
The Cas~A echoes represent the first indirect $''$view$''$ of a shock breakout via the thermal dust emission from echoing clouds.  
 
ED acknowledges the support of NASA's LTSA03-0000-065. We thank the referee for his/her critical comments. 
This work is based on observations made with the Spitzer Space Telescope, which is operated by the Jet Propulsion Laboratory, California Institute of Technology, under contract with NASA.


\newpage

\newpage

\begin{deluxetable}{lcccc}
\tablewidth{0pt}
\tablecaption{Observed Properties of Echoing Clouds}
\tablehead{
\colhead{Target} &
 \colhead{ Location (RA: DEC)} &
 \colhead{ $\alpha$ (\arcsec)\tablenotemark{1}} &
 \colhead{$r_{50}$ (lyr)\tablenotemark{2}} &
 \colhead{$r_{320}$ (lyr)\tablenotemark{3}} 
  }
 \startdata
Echo 1 & 23h23\arcmin38.57\arcsec: \ +58d52\arcmin54.4\arcsec  &  259\arcsec & 27.0 & 160.3 \\
Echo 2 & 23h22\arcmin50.58\arcsec:  +58d47\arcmin30.6\arcsec &    300\arcsec & 27.6 & 160.4 \\
Echo 3 & 23h22\arcmin 41.55\arcsec: \ +58d48\arcmin37.0\arcsec  &  359\arcsec & 28.7 & 160.6\\
Echo 4 & 23h22\arcmin50.98\arcsec: \ +58d41\arcmin49.4\arcsec  &  508\arcsec & 32.5 & 161.2\\
Echo 5 & 23h24\arcmin12.37\arcsec: \ +59d01\arcmin12.8\arcsec  &  820\arcsec & 44.5 & 163.1\\
Echo 6 & 23h21\arcmin39.96\arcsec:  +59d34\arcmin26.9\arcsec  &  2860\arcsec & 258 & 198\\ 
Echo 7 & 23h26\arcmin38.62\arcsec:  +57d01\arcmin19.4\arcsec  &    6630\arcsec & 1172 & 359\\
 \enddata
\tablenotetext{1}{Angular distances were calculated from the optical expansion center located at (RA: DEC) = (23h23\arcmin27.77\arcsec:  +58d48m49.4s)  \citep{Thorstensen01} }
\tablenotetext{2}{Echo distances from the SN for a delay time of $t=50$~yr.}
\tablenotetext{3}{Echo distances from the SN for a delay time of $t=320$~yr.}
\end{deluxetable}


\begin{deluxetable}{llllllll}
\tabletypesize{\footnotesize}
\tablewidth{0pt}
\tablecaption{Derived Model Properties\tablenotemark{1}}
\tablehead{
\colhead{Echo \#} &
 \colhead{ } &
 \colhead{1} &
  \colhead{2} &
   \colhead{3} &
    \colhead{4} &
     \colhead{5} &
      \colhead{6} 
   }
 \startdata
\olc: & \xsi\ & 0.6 & 0.6 & 0.8 & 0.2 & 0.3 & 0.6  \\
 & $L_b$  &0.60 & 0.60 & 0.81 & 0.20 & 0.31 & 0.92  \\
  & $N_H$  &2.7 & 1.6 & 1.9 & 1.3 & 3.0 & 9.7  \\
  & $\chi^2$ &2.20 & 2.46 & 4.62 & 1.81 & 1.72 & 1.31 \\ 
\hline \\
\euva: & \xsi\ & 0.2 & 0.2 & 0.2 & 0.04 & 0.08 & 0.1  \\
 & $L_b$  &0.20 & 0.20 & 0.20 & 0.041 & 0.083 & 0.15  \\
   & $N_H$  & 4.7 & 2.8 & 4.2 & 4.4 & 7.1 & 32.0   \\
    & $\chi^2$ & 1.67 & 1.66 & 3.98 & {\bf 1.00} & 1.03 & 1.13 \\
 \euvb: & \xsi\ & 0.2 & 0.2 & 0.4 & 0.06 & 0.1 & 0.2  \\
 & $L_b$  &0.20 & 0.20 & 0.40 & 0.06 & 0.10 & 0.31  \\
   & $N_H$  &7.7 & 4.6 & 3.5 & 4.9 & 9.5 & 26.9   \\
    & $\chi^2$ &2.40 & 2.94 & 3.21 & 1.05 & 1.40 & {\bf 1.00} \\
\euvc: & \xsi\ & 2.0 & 2.0 & 2.0 & 0.4 & 0.8 & 1.0  \\
 & $L_b$  &2.0 & 2.0 & 2.0 & 0.41 & 0.83 & 1.53  \\
   & $N_H$  &3.5 & 2.1 & 3.1 & 3.2 & 5.2 & 23.8  \\
    & $\chi^2$ &1.93 & 1.74 & 3.16 & 1.03 & {\bf 1.00} & 1.10 \\
 \hline \\
\uva: & \xsi\ & 0.1 & 0.1 & 0.2 & 0.02 & 0.06 & 0.1  \\
 & $L_b$  &0.10 & 0.10 & 0.20 & 0.020 & 0.062 & 0.15   \\
   & $N_H$  &5.8 & 3.5 & 2.8 & 5.4 & 5.9 & 20.6  \\
    & $\chi^2$ &{\bf 1.00} & 1.52 & {\bf 1.00} & 1.64 & 1.10 & 1.09 \\
\uvb: & \xsi\ & 0.2 & 0.2 & 0.2 & 0.04 & 0.08 & 0.1  \\
 & $L_b$  &0.20 & 0.20 & 0.20 & 0.041 & 0.083 & 0.15   \\
   & $N_H$  &3.6 & 2.2 & 3.2 & 3.2 & 5.3 & 24.0   \\
    & $\chi^2$ &1.14 & {\bf 1.00} & 1.45 & 1.28 & 1.07 & 1.29 \\
\uvc: & \xsi\ & 0.4 & 0.2 & 0.4 & 0.06 & 0.1 & 0.2  \\
 & $L_b$  &0.40 & 0.20 & 0.40 & 0.061 & 0.10 & 0.31   \\
   & $N_H$  &3.0 & 3.4 & 2.7 & 3.5 & 6.9 & 20.0   \\
    & $\chi^2$ &1.50 & 2.23 & 1.17 & 1.41 & 1.61 & 1.22 \\
 \enddata
\tablenotetext{1}{\xsi\ is the value of $\xi$ [see eq. (\ref{flux})] that gives the best fitting incident flux [eq. (\ref{lburst})]. Burst luminosities are in units of $10^{12}$~\lsun, and echo column densities are in units of $10^{17}$~cm$^{-2}$. For each echo, $\chi^2$ is renormalized to 1.00 for the best fitting model.}
\end{deluxetable}


\begin{deluxetable}{ccc}
\tabletypesize{\footnotesize}
\tablewidth{0pt}
\tablecaption{Average Burst Luminosities\tablenotemark{1}}
\tablehead{
\colhead{Burst Model} &
       \colhead{$<L_b>$} &
       \colhead{$\sigma$} \\
 \colhead{ }&
 \colhead{($10^{12}$~\lsun)}&
 \colhead{($10^{12}$~\lsun)}
  }
 \startdata
 \olc& 0.57 & 0.28 \\
 \hline
 \euva &  0.15 & 0.070 \\
\euvb &  0.21 & 0.13 \\
\euvc &  1.47 & 0.70  \\
\hline 
\uva &  0.11 & 0.064  \\
\uvb &  0.15 & 0.07  \\
\uvc &  0.25 & 0.15  \\
  \enddata
\tablenotetext{1}{$<L_b>$ is the value of $L_b$ averaged over all echoing  clouds for each burst model, and $\sigma$ is the r.m.s. dispersion of the value.}
\end{deluxetable}


\begin{deluxetable}{ccc}
\tabletypesize{\footnotesize}
\tablewidth{0pt}
\tablecaption{Average Echo Column Densities\tablenotemark{1}}
\tablehead{
\colhead{Echo \#} &
       \colhead{$<N_H>$} &
       \colhead{$\sigma$} \\
 \colhead{ }&
 \colhead{($10^{17}$~cm$^{-2}$)}&
 \colhead{($10^{17}$~cm$^{-2}$)}
  }
 \startdata
 1& 4.7 &       1.8 \\
 2 &    3.1 &      0.9 \\
  3 &  3.3 &      0.6 \\
  4 &    4.1 &      1.0 \\
   5 &    6.7 &       1.6 \\
   6&   24.5 &       4.4  \\

  \enddata
\tablenotetext{1}{$<N_H>$ is the value of $N_H$ averaged over all EUV-UV burst models for each echo, and $\sigma$ is the r.m.s. dispersion of the value.}
\end{deluxetable}

\clearpage

\clearpage

\begin{thebibliography}{33}
\expandafter\ifx\csname natexlab\endcsname\relax\def\natexlab#1{#1}\fi

\bibitem[{{Bieging} {et~al.}(1991){Bieging}, {Goss}, \& {Wilcots}}]{bieging91}
{Bieging}, J.~H., {Goss}, W.~M., \& {Wilcots}, E.~M. 1991, \apjs, 75, 999

\bibitem[{{Blinnikov} {et~al.}(2000){Blinnikov}, {Lundqvist}, {Bartunov},
  {Nomoto}, \& {Iwamoto}}]{blinnikov00}
{Blinnikov}, S., {Lundqvist}, P., {Bartunov}, O., {Nomoto}, K., \& {Iwamoto},
  K. 2000, \apj, 532, 1132

\bibitem[{{Bode} {et~al.}(2004){Bode}, {O'Brien}, \& {Simpson}}]{bode04}
{Bode}, M.~F., {O'Brien}, T.~J., \& {Simpson}, M. 2004, \apjl, 600, L63

\bibitem[{{Bond} {et~al.}(2003){Bond}, {Henden}, {Levay}, {Panagia}, {Sparks},
  {Starrfield}, {Wagner}, {Corradi}, \& {Munari}}]{bond03}
{Bond}, H.~E., {Henden}, A., {Levay}, Z.~G., {et~al.} 2003, \nat, 422, 405

\bibitem[{Brown {et~al.}(2008)}]{brown08}
Brown, P.~J. {et~al.} 2008, \apj, submitted, arXiv:0803.1265

\bibitem[{{Crotts}(1988)}]{crotts88}
{Crotts}, A.~P.~S. 1988, \apjl, 333, L51

\bibitem[{{Draine} \& {Li}(2007)}]{draine07}
{Draine}, B.~T. \& {Li}, A. 2007, \apj, 657, 810

\bibitem[{{Dwek} \& {Felten}(1992)}]{dwek92d}
{Dwek}, E. \& {Felten}, J.~E. 1992, \apj, 387, 551

\bibitem[{{Ennis} {et~al.}(2006){Ennis}, {Rudnick}, {Reach}, {Smith}, {Rho},
  {DeLaney}, {Gomez}, \& {Kozasa}}]{ennis06}
{Ennis}, J.~A., {Rudnick}, L., {Reach}, W.~T., {et~al.} 2006, \apj, 652, 376

\bibitem[{{Ensman} \& {Burrows}(1992)}]{ensman92}
{Ensman}, L. \& {Burrows}, A. 1992, \apj, 393, 742

\bibitem[{{Falk}(1978)}]{falk78}
{Falk}, S.~W. 1978, \apjl, 225, L133

\bibitem[{{Fransson} {et~al.}(1989){Fransson}, {Cassatella}, {Gilmozzi},
  {Kirshner}, {Panagia}, {Sonneborn}, \& {Wamsteker}}]{fransson89a}
{Fransson}, C., {Cassatella}, A., {Gilmozzi}, R., {et~al.} 1989, \apj, 336, 429

\bibitem[{{Fransson} \& {Lundqvist}(1989)}]{fransson89b}
{Fransson}, C. \& {Lundqvist}, P. 1989, \apjl, 341, L59

\bibitem[{{Hines} {et~al.}(2004){Hines}, {Rieke}, {Gordon}, {Rho}, {Misselt},
  {Woodward}, {Werner}, {Krause}, {Latter}, {Engelbracht}, {Egami}, {Kelly},
  {Muzerolle}, {Stansberry}, {Su}, {Morrison}, {Young}, {Noriega-Crespo},
  {Padgett}, {Gehrz}, {Polomski}, {Beeman}, \& {Haller}}]{hines04}
{Hines}, D.~C., {Rieke}, G.~H., {Gordon}, K.~D., {et~al.} 2004, \apjs, 154, 290

\bibitem[{{Kapteyn}(1902)}]{kapteyn02}
{Kapteyn}, J.~C. 1902, Astronomische Nachrichten, 157, 201

\bibitem[{{Kim} {et~al.}(2008){Kim}, {Rieke}, {Krause}, {Misselt},
  {Indebetouw}, \& {Johnson}}]{kim08}
{Kim}, Y., {Rieke}, G.~H., {Krause}, O., {et~al.} 2008, \apj, 678, 287

\bibitem[{{Klein} \& {Chevalier}(1978)}]{klein78}
{Klein}, R.~I. \& {Chevalier}, R.~A. 1978, \apjl, 223, L109

\bibitem[{{Krause} {et~al.}(2005){Krause}, {Rieke}, {Birkmann}, {Le Floc'h},
  {Gordon}, {Egami}, {Bieging}, {Hughes}, {Young}, {Hinz}, {Quanz}, \&
  {Hines}}]{krause05}
{Krause}, O., {Rieke}, G.~H., {Birkmann}, S.~M., {et~al.} 2005, Science, 308,
  1604

\bibitem[{{Mathis} {et~al.}(1983){Mathis}, {Mezger}, \& {Panagia}}]{mathis83}
{Mathis}, J.~S., {Mezger}, P.~G., \& {Panagia}, N. 1983, \aap, 128, 212

\bibitem[{{Matzner} \& {McKee}(1999)}]{matzner99}
{Matzner}, C.~D. \& {McKee}, C.~F. 1999, \apj, 510, 379

\bibitem[{{Panagia} {et~al.}(1991){Panagia}, {Gilmozzi}, {Macchetto}, {Adorf},
  \& {Kirshner}}]{panagia91}
{Panagia}, N., {Gilmozzi}, R., {Macchetto}, F., {Adorf}, H.-M., \& {Kirshner},
  R.~P. 1991, \apjl, 380, L23

\bibitem[{{Reed} {et~al.}(1995){Reed}, {Hester}, {Fabian}, \&
  {Winkler}}]{reed95}
{Reed}, J.~E., {Hester}, J.~J., {Fabian}, A.~C., \& {Winkler}, P.~F. 1995,
  \apj, 440, 706

\bibitem[{{Rest} {et~al.}(2005){Rest}, {Suntzeff}, {Olsen}, {Prieto}, {Smith},
  {Welch}, {Becker}, {Bergmann}, {Clocchiatti}, {Cook}, {Garg}, {Huber},
  {Miknaitis}, {Minniti}, {Nikolaev}, \& {Stubbs}}]{rest05}
{Rest}, A., {Suntzeff}, N.~B., {Olsen}, K., {et~al.} 2005, \nat, 438, 1132

\bibitem[{{Rho} {et~al.}(2008){Rho}, {Kozasa}, {Reach}, {Smith}, {Rudnick},
  {DeLaney}, {Ennis}, {Gomez}, \& {Tappe}}]{rho08}
{Rho}, J., {Kozasa}, T., {Reach}, W.~T., {et~al.} 2008, \apj, 673, 271

\bibitem[{{Soderberg} {et~al.}(2008){Soderberg}, {Berger}, {Page}, {Schady},
  {Parrent}, {Pooley}, {Wang}, {Ofek}, {Cucchiara}, {Rau}, {Waxman}, {Simon},
  {Bock}, {Milne}, {Page}, {Barthelmy}, {Beardmore}, {Bietenholz}, {Brown},
  {Burrows}, {Burrows}, {Byrngelson}, {Cenko}, {Chandra}, {Cummings}, {Fox},
  {Gal-Yam}, {Gehrels}, {Immler}, {Kasliwal}, {Kong}, {Krimm}, {Kulkarni},
  {Meszaros}, {Nakar}, {O'Brien}, {de Pasquale}, {Racusin}, \&
  {Rea}}]{soderberg08}
{Soderberg}, A.~M., {Berger}, E., {Page}, K., {et~al.} 2008, ArXiv e-prints,
  0802.1712

\bibitem[{{Sonneborn} {et~al.}(1997){Sonneborn}, {Fransson}, {Lundqvist},
  {Cassatella}, {Gilmozzi}, {Kirshner}, {Panagia}, \&
  {Wamsteker}}]{sonneborn97}
{Sonneborn}, G., {Fransson}, C., {Lundqvist}, P., {et~al.} 1997, \apj, 477, 848

\bibitem[{{Thorstensen} {et~al.}(2001){Thorstensen}, {Fesen}, \& {van den
  Bergh}}]{Thorstensen01}
{Thorstensen}, J.~R., {Fesen}, R.~A., \& {van den Bergh}, S. 2001, \aj, 122,
  297

\bibitem[{{Ungerechts} {et~al.}(2000){Ungerechts}, {Umbanhowar}, \&
  {Thaddeus}}]{ungerechts00}
{Ungerechts}, H., {Umbanhowar}, P., \& {Thaddeus}, P. 2000, \apj, 537, 221

\bibitem[{{van den Bergh}(1965)}]{van-den-bergh65}
{van den Bergh}, S. 1965, \pasp, 77, 269

\bibitem[{{Wolfire} {et~al.}(2003){Wolfire}, {McKee}, {Hollenbach}, \&
  {Tielens}}]{wolfire03}
{Wolfire}, M.~G., {McKee}, C.~F., {Hollenbach}, D., \& {Tielens}, A.~G.~G.~M.
  2003, \apj, 587, 278

\bibitem[{{Woosley}(1988)}]{woosley88}
{Woosley}, S.~E. 1988, \apj, 330, 218

\bibitem[{{Young} {et~al.}(2006){Young}, {Fryer}, {Hungerford}, {Arnett},
  {Rockefeller}, {Timmes}, {Voit}, {Meakin}, \& {Eriksen}}]{young06}
{Young}, P.~A., {Fryer}, C.~L., {Hungerford}, A., {et~al.} 2006, \apj, 640, 891

\bibitem[{{Zubko} {et~al.}(2004){Zubko}, {Dwek}, \& {Arendt}}]{zubko04}
{Zubko}, V., {Dwek}, E., \& {Arendt}, R.~G. 2004, \apjs, 152, 211

\end{thebibliography}
\end{document}